\documentclass[manuscript,screen,review=false]{acmart}

\usepackage{flushend}
\usepackage{balance}
\usepackage{multirow}
\usepackage{booktabs}
\usepackage{graphicx}
\usepackage{subfigure} 
\usepackage{url}       
\usepackage{stfloats}  
\usepackage{amsmath}   
\usepackage{array}
\usepackage{breqn}
\usepackage{amsfonts}
\usepackage{courier}

\usepackage{amssymb}
\usepackage{balance}
\usepackage{multirow,tabularx}
\usepackage{longtable}
\usepackage{booktabs,longtable}
\usepackage{hhline}
\usepackage{dsfont}
\usepackage{fancyhdr}
\usepackage{epstopdf}
\usepackage{algorithm}
\usepackage{enumitem}
\usepackage{algorithmic}
\usepackage[bottom]{footmisc}
\usepackage{multicol}
\usepackage{multirow}
\usepackage{verbatim}
\usepackage{comment}
\usepackage{longtable}
\usepackage{arydshln}
\usepackage{color} 
\usepackage{xcolor}
\usepackage{caption}
\usepackage{makecell}

\usepackage[most]{tcolorbox}

\AtBeginDocument{%
  }

\setcopyright{acmlicensed}
\copyrightyear{2024}
\acmYear{2024}
\acmJournal{TOIS}
\acmDOI{XXXXXXX.XXXXXXX}

\acmISBN{978-1-4503-XXXX-X/18/06}

\begin{document}

\title{A Comprehensive Survey on Composed Image Retrieval}

\author{Xuemeng Song}
\orcid{0000-0002-5274-4197}
\affiliation{%
    \institution{Shandong University}
    \city{Qingdao}
   \country{China}
}
\email{sxmustc@gmail.com}

\author{Haoqiang Lin}
\orcid{0009-0000-5768-5467}
\affiliation{%
    \institution{Shandong University}
    \city{Qingdao}
   \country{China}
}
\email{zichaohq@gmail.com}

\author{Haokun Wen}
\orcid{0000-0003-0633-3722}
\affiliation{%
    \institution{\mbox{Harbin Institute of Technology~(Shenzhen) }}
    \city{Shenzhen}
    \country{China}
}
\affiliation{%
    \institution{City University of Hong Kong}
    \city{Hong Kong}
    \country{China}
}
\email{whenhaokun@gmail.com}

 \author{Bohan Hou}
 \orcid{0009-0008-0751-166X}
 \affiliation{%
    \institution{Shandong University}
     \city{Qingdao}
    \country{China}
 }
  \email{bohanhou@foxmail.com}

  \author{Mingzhu Xu}
 \orcid{0000-0002-1492-0970}
 \affiliation{%
    \institution{Shandong University}
     \city{JiNan}
    \country{China}
 }
  \email{xumingzhu@sdu.edu.cn}

\author{Liqiang Nie}
\orcid{0000-0003-1476-0273}
\affiliation{%
	\institution{\mbox{Harbin Institute of Technology~(Shenzhen)}}
	 \city{Shenzhen}
     \country{China}
}
\email{nieliqiang@gmail.com}

\renewcommand{\shortauthors}{Xuemeng Song et al.}
\thanks{This work has been submitted to the ACM for possible publication. Copyright may be transferred without notice, after which this version may no longer be accessible.}

\begin{abstract}
Composed Image Retrieval (CIR) is an emerging yet challenging task that allows users to search for target images using a multimodal query, comprising a reference image and a modification text specifying the user's desired changes to the reference image. Given its significant academic and practical value, CIR has become a rapidly growing area of interest in the computer vision and machine learning communities, particularly with the advances in deep learning. To the best of our knowledge, there is currently no comprehensive review of CIR to provide a timely overview of this field. Therefore,  we synthesize insights from over 120 publications in top conferences and journals, including ACM TOIS, SIGIR, and CVPR.
In particular, we systematically categorize existing supervised CIR and zero-shot CIR models using a fine-grained taxonomy. For a comprehensive review, we also briefly discuss approaches for tasks closely related to CIR, such as attribute-based CIR and dialog-based CIR. Additionally, we summarize benchmark datasets for evaluation and analyze existing supervised and zero-shot CIR methods by comparing experimental results across multiple datasets. Furthermore, we present promising future directions in this field, offering practical insights for researchers interested in further exploration. The curated collection of related works is maintained and continuously updated in \href{https://github.com/haokunwen/Awesome-Composed-Image-Retrieval}{\textsc{Awesome-CIR}} repository.
\end{abstract}



\received{20 February 2007}
\received[revised]{12 March 2009}
\received[accepted]{5 June 2009}

\maketitle

\label{sec: intro}
\section{Introduction}

Image retrieval has been a fundamental task in computer vision and database management since the 1970s~\cite{datta2008image}, serving as a cornerstone for various applications, such as face recognition~\cite{fu2021dvg}, fashion retrieval~\cite{yang2020generative}, and person re-identification~\cite{li2019pose}. Traditional image retrieval systems primarily rely on unimodal queries, using either text or images to convey a user's search intent~\cite{chua1994content, qu2021dynamic, qu2020context, rao2022does}. However, users often struggle to clearly express their search intent through a single text query or to find the perfect image that accurately represents it.
To address these limitations and provide greater flexibility, composed image retrieval~(CIR)~\cite{vo2019tirg} emerged in 2019, which allows users to express their search intent by a reference image combined with a textual description specifying the desired modifications. By enabling users to utilize more nuanced search queries, CIR offers significant potential to enhance search experiences across domains, such as e-commerce~\cite{du2023multi} and internet search engines~\cite{jandial2022sac,wu2021fiq,weicom, shf}.

\begin{figure}[!t]
		\centering
		\includegraphics[scale=0.55]{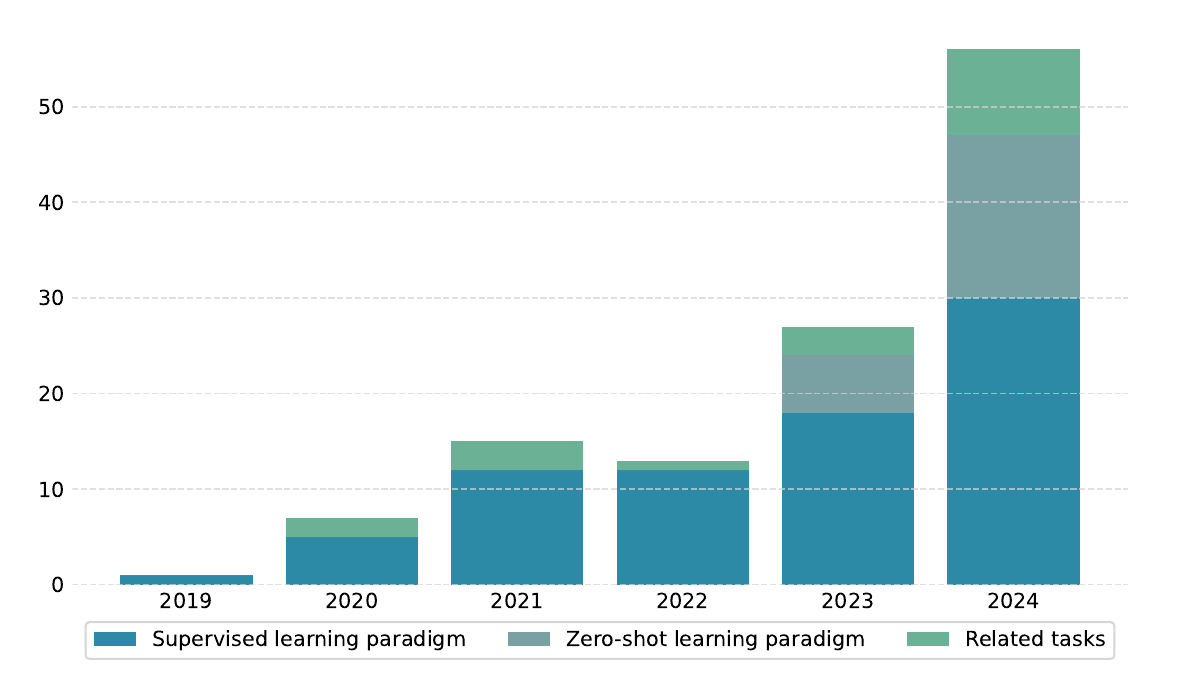}
         
	    \caption{The number of papers trending on the task of composed image retrieval and its related task since 2019.}\label{paper_num}
\end{figure}

The concept of CIR, which allows users to utilize a multimodal query to express their search intent, can be easily adapted for various real-world retrieval scenarios. For example, the reference image could be replaced with a reference video to enable composed video retrieval, or single-turn CIR could evolve into dialog-based multi-turn image retrieval. Since its introduction in 2019, CIR has garnered increasing research attention due to its potential value across various domains. As illustrated in Figure~\ref{paper_num}, the number of publications on CIR is increasing rapidly. 
To summarize the past and current achievements in this rapidly developing field, we present a comprehensive overview of work conducted up to November 2024. Existing studies primarily focus on addressing the following key challenges.
1) \textbf{Multimodal Query Fusion.} In CIR, the modification text and reference image play complementary roles in conveying the user's search intent. The modification text typically specifies changes in certain attributes of the reference image. For instance, given the modification requirement, ``I want the dress to be black and more professional'', only the color and style of the dress in the reference image should be changed, while other attributes of the reference image should be kept unchanged. Due to this nature, how to achieve an effective multimodal fusion for accurately comprehending the multimodal query poses the first challenge.
2) \textbf{Target Images Matching.} The semantic gap between the multimodal query and target images presents a significant challenge due to their heterogeneous representations. Additionally, the brevity of modification texts can lead to ambiguity. For example, the text ``I want to change the dress to longer sleeves and yellow in color'' could have multiple interpretations: the sleeves could change from sleeveless to either short or long, and the color could range from light to dark yellow. Such ambiguity suggests that multiple target images could satisfy the given query. Therefore, bridging this semantic gap and managing the one-to-many query-to-target matching relationship is crucial for accurate query-target matching.
3) \textbf{Scale of Training Data}. Training CIR models typically requires triplets in the form of <reference image, modification text, target image>. For each triplet, the reference-target image pair is often generated using a heuristic strategy, while the modification text is usually annotated by humans. Creating such training samples is both costly and labor-intensive, which significantly restricts the size of benchmark datasets. Consequently, addressing the issue of insufficient training data to improve the model's generalization capabilities remains a significant challenge.

Existing work in this area can be broadly divided into two main categories: supervised learning-based approaches and zero-shot learning-based approaches. The key distinction between these methods lies in the availability of annotated training triplets. Supervised approaches rely on annotated triplets from the dataset to train the model, while zero-shot approaches leverage large-scale, easily accessible data, such as image-text pairs, for pre-training without requiring annotated triplets for optimization.
To facilitate deeper analysis, we establish a fine-grained taxonomy for each category. For supervised CIR approaches, we summarize existing methods based on the four key components of the general framework: feature extraction, image-text fusion, target matching, and data augmentation. For zero-shot composed image retrieval (ZS-CIR) approaches, we classify methods into three groups: textual-inversion-based, pseudo-triplet-based, and training-free.
As previously mentioned, the concept of using a composed multimodal query can be adapted for various scenarios. Beyond the primary task of CIR, several related tasks also involve composed queries, such as reference image plus attribute manipulation, sketch plus modification text, and video plus modification text. Since these tasks are closely related to CIR, we include their recent advancements to provide a comprehensive review of the topic. Based on the type of multimodal query, we categorize these related tasks into five groups: attribute-based, sketch-based, remote sensing-based, dialog-based, and video-based.
\begin{figure}[!t]
		\centering
		\includegraphics[scale=0.45]{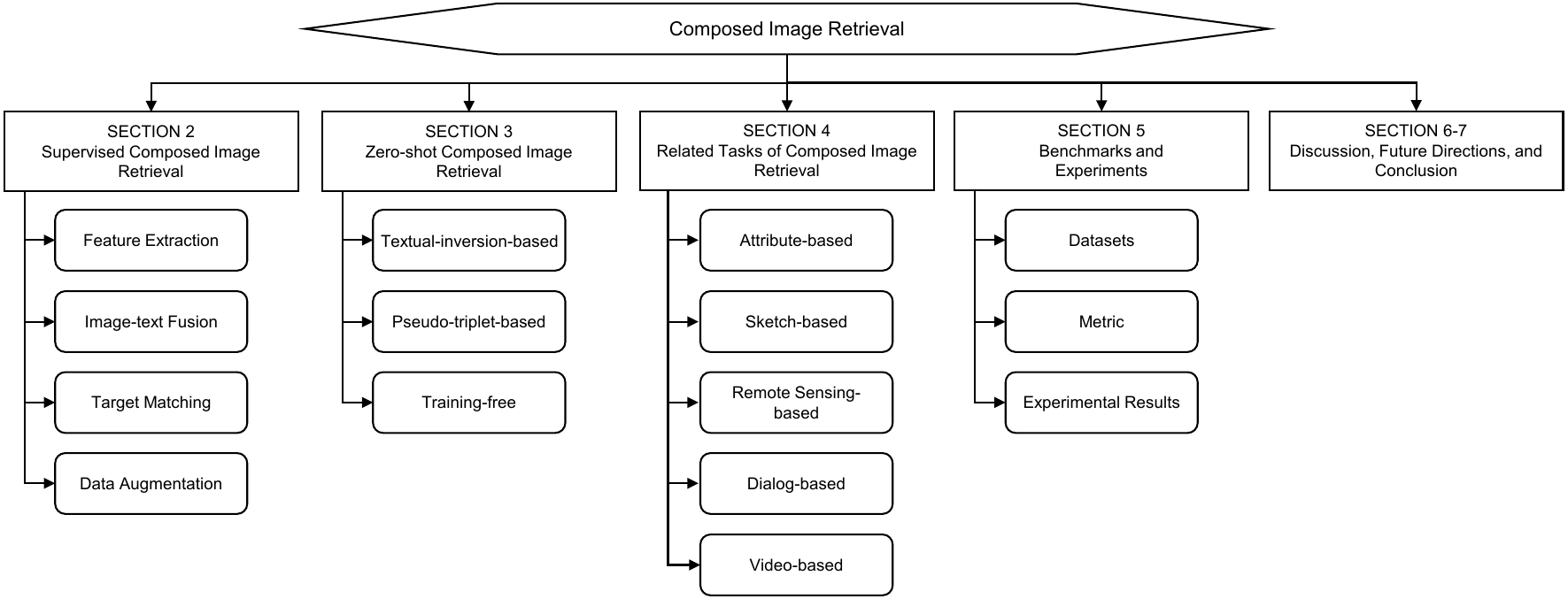}
	    \caption{Organization of the present survey.}\label{organization}
           
\end{figure}

In summary, our main contributions are as follows:
\begin{itemize}[leftmargin=20pt]
	\item To the best of our knowledge, this paper presents the first comprehensive review of CIR, incorporating over 120 primary studies. It aims to provide a timely and insightful overview to guide future research in this rapidly advancing field.
        \item We systematically organize research findings, technical approaches, benchmarks, and experiments to deepen the understanding of this field. Additionally, we propose an elaborate taxonomy of methods, catering to the diverse needs of readers.
        \item CIR remains an emerging area of research. Based on the surveyed literature, we identify several key research challenges and propose potential future directions, offering forward-looking guidance for researchers in this domain.
\end{itemize}

The remainder of this paper is organized as depicted in Figure~\ref{organization}. Sections~$2$ and $3$ review supervised CIR models and zero-shot CIR models, respectively. Section~$4$ introduces  tasks related to CIR. Section~$5$ describes the currently available datasets, evaluation metrics used, and experimental results from existing approaches. Finally, we discuss possible future research directions in Section~$6$ and conclude the work in Section~$7$. 
\label{sec: meth_q}

\begin{figure}[!t]
		\centering
		\includegraphics[scale=0.50]{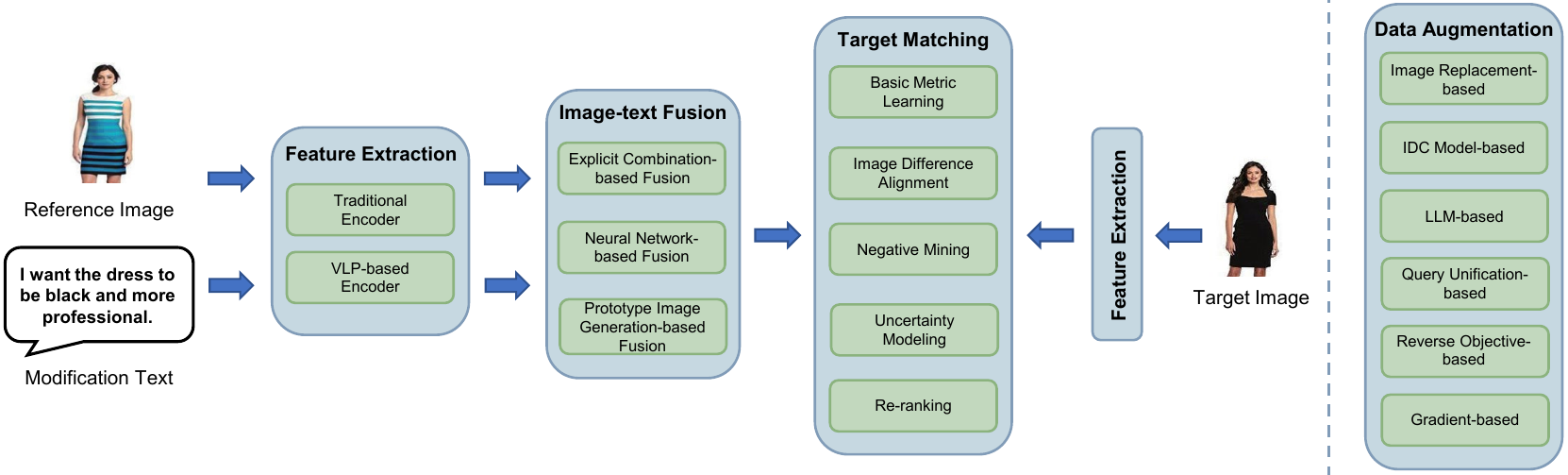}
	    \caption{The illustration of the standard framework of supervised composed image retrieval.}\label{framework_scir}
\end{figure}

\begin{table*}
\scriptsize
\centering
 \caption{\textbf{Summarization of main supervised composed image retrieval approaches.}}
 \resizebox{14.5cm}{!}{
\begin{tabular}{|c|c|c|c|cc|c|}
\hline
\multicolumn{2}{|c|}{Fusion Strategy} & Method & Year & Image Encoder & Text Encoder & Other Aspect\\ 
\hline


\multirow{22}{*}{ \makecell{Explicit Combination-based \\ Fusion}  } & \multirow{18}{*}{ \makecell{Transformed \\ image-and-Residual}} & TIRG~\cite{vo2019tirg} & 2019 & ResNet-17 & LSTM & CSS Dataset \\ \cline{3-7}
&  & VAL~\cite{chen2020val} & 2020 & ResNet-50, MobileNet & LSTM & Hierarchical Matching\\ \cline{3-7}
&  & JVSM~\cite{chen2020jvsm} & 2020 & MobileNet & LSTM & Joint Visual Semantic Matching \\ \cline{3-7}
&  & DATIR~\cite{gu2021datir} & 2021 & ResNet-50, MobileNet & LSTM & Hierarchical Matching \\ \cline{3-7}
&  & DCNet~\cite{kim2021dcnet} & 2021 & ResNet-50 & GloVe+MLP & Image Difference Alignment \\ \cline{3-7}
&  & MGF~\cite{liu2021mgf} & 2021 & ResNet-17 & LSTM & Online Groups Matching \\ \cline{3-7}
&  & MCR~\cite{zhang2021mcr} & 2021 & ResNet-50 & LSTM & Image Difference Alignment \\ \cline{3-7}
&  & CLVC-Net~\cite{wen2021clvcnet} & 2021 & ResNet-50 & LSTM & Mutual Enhancement \\ \cline{3-7}
&  & SAC~\cite{jandial2022sac} & 2022 & ResNet-50 & BERT & Matching Optimization \\ \cline{3-7}
&  & EER~\cite{zhang2022eer} & 2022 & ResNet-50 & LSTM & Semantic Space Alignment \\ \cline{3-7}
&  & CRN~\cite{yang2023crn} & 2023 & Swin Transformer & LSTM & Cross Relation Retrieval \\ \cline{3-7}
&  & MLCLSAP~\cite{zhang2023MLCLSAP} & 2023 & ResNet-50 & LSTM & Image Difference Alignment \\ \cline{3-7}
&  & TG-CIR~\cite{wen2023tgcir} & 2023 & CLIP-B & CLIP-B & Target Similarity Guidance \\ \cline{3-7}
&  & Ranking-aware~\cite{chen2023ranking} & 2023 & CLIP(RN50) & CLIP(RN50) & Uncertainty Modeling \\ \cline{3-7}
&  & MCEM~\cite{zhang2024mcem} & 2024 & ResNet-18, ResNet-50 & LSTM & Negative Example Mining \\ \cline{3-7}
&  & DWC~\cite{huang2024dwc} & 2024 & ResNet-50, CLIP(RN50) & LSTM, CLIP(RN50) & Mutual Enhancement \\ \cline{3-7}
&  & Css-Net~\cite{zhang2024cssnet} & 2024 & ResNet-18, ResNet-50 & RoBERTa & Collaborative Matching \\ \cline{3-7}
&  & AlRet~\cite{xu2024alret} & 2024 & ResNet-50, CLIP(RN50) & LSTM, CLIP(RN50) & Composition and Decomposition \\ \cline{2-7}

&  \multirow{4}{*}{Content-and-Style} & CoSMo~\cite{lee2021cosmo} & 2021 & ResNet-18, ResNet-50 & LSTM & - \\ \cline{3-7}
&  & LSC4TCIR~\cite{chawla2021lsc4cir} & 2021 & ResNet-50 & GRU & - \\ \cline{3-7}
&  & PCaSM~\cite{zhang2023pcasm} & 2023 & ResNet-18, ResNet-50 & LSTM & - \\ \cline{3-7}
&  & SPIRIT~\cite{chen2024spirit} & 2024 & CLIP(RN50x4) & CLIP(RN50x4) & Patch-level Graph Reasoning \\  \cline{2-7}

\hline

\multirow{47}{*}{ \makecell{Neural Network-based \\ Fusion}  }& \multirow{18}{*}{MLP-based} & ComposeAE~\cite{anwaar2021ComposeAE} & 2021 & ResNet-17 & BERT & Rotational Symmetry \\ \cline{3-7}
&  & Combiner~\cite{baldrati2022Combiner} & 2022 & CLIP(RN50x4) & CLIP(RN50x4) & - \\ \cline{3-7}
&  & CLIP4CIR~\cite{baldrati2022CLIP4CIR} & 2022 & CLIP(RN50x4) & CLIP(RN50x4) & Fine-tune Strategy \\ \cline{3-7}
&  & PL4CIR~\cite{zhao2022PL4CIR} & 2022 & CLIP(RN50), CLIP-B & CLIP(RN50), CLIP-B & Fashion-based Fine-tuning \\ \cline{3-7}
&  & ARTEMIS~\cite{delmas2022artemis} & 2022 & ResNet-18, ResNet-50 & Bi-GRU, LSTM & - \\ \cline{3-7}
&  & CLIP4CIR2~\cite{baldrati2023CLIP4CIR2} & 2023 & CLIP(RN50x4) & CLIP(RN50x4) & Fine-tune Strategy \\ \cline{3-7}
&  & DSCN~\cite{li2023dscn} & 2023 & ResNet-18, ResNet-50 & Bi-GRU & Hierarchical Matching \\ \cline{3-7}
&  & CLIP-CD~\cite{lin2023clip_cd} & 2023 & CLIP(RN50x4) & CLIP(RN50x4) & Pseudo Triplet Generation \\ \cline{3-7}
&  & BLIP4CIR~\cite{liu2024blip4cir} & 2024 & BLIP & BLIP & Reverse Learning \\ \cline{3-7}
&  & CMAP~\cite{li2024cmap} & 2024 & ResNet-50 & Bi-GRU, LSTM & Hierarchical Matching \\ \cline{3-7}
&  & CAFF~\cite{wan2024caff} & 2024 & CLIP(RN50) & CLIP(RN50) & Fashion-based Fine-tuning \\ \cline{3-7}
&  & MANME~\cite{li2023manme} & 2024 & ResNet-50 & Bi-GRU, LSTM & Hierarchical Matching \\ \cline{3-7}
&  & NSFSE~\cite{wang2024NSFSE} & 2024 & ResNet-50, ResNet-152 & Bi-GRU, LSTM & Negative Sensitive Framework \\ \cline{3-7}
&  & SHAF~\cite{yan2024shaf} & 2024 & FashionCLIP & FashionCLIP & Hierarchical Alignment \\ \cline{3-7}
&  & CLIP-ProbCR~\cite{li2024clip} & 2024 & CLIP & CLIP & Uncertainty Modeling \\ \cline{3-7}
&  & DMOT~\cite{dmot} & 2024 & BLIP & BLIP & - \\ \cline{3-7}
&  & DQU-CIR~\cite{wen2024dqu} & 2024 & CLIP-H & CLIP-H & Data Augmentation \\ \cline{3-7}
&  & SADN~\cite{wang2024sadn} & 2024 & CLIP(RN50x4) & CLIP(RN50x4) & Neighborhood Distillation \\ \cline{2-7}

& \multirow{10}{*}{Cross-attention-based} & LBF~\cite{hosseinzadeh2020lbf} & 2020 & Faster R-CNN & TEP & Coarse and Fine Retrieval \\ \cline{3-7}
&  & MAAF~\cite{dodds2020maaf} & 2020 & ResNet-50 & LSTM &- \\ \cline{3-7}
&  & ProVLA~\cite{hu2023provla} & 2023 & Swin Transformer& BERT & Negative Example Mining \\ \cline{3-7}
&  & ComqueryFormer~\cite{xu2023ComqueryFormer} & 2023 & Swin Transformer & BERT & Hierarchical Matching \\ \cline{3-7}
&  & LGLI~\cite{huang2023lgli} & 2023 & ResNet-18 & LSTM & - \\ \cline{3-7}
&  & ACNet~\cite{li2023acnet} & 2023 & ResNet-50 & Bi-GRU & Image Difference Alignment \\ \cline{3-7}
&  & Re-ranking~\cite{liu2023rerank} & 2024 & BLIP & BLIP & Re-rank \\ \cline{3-7}
&  & CASE~\cite{levy2024case} & 2024 & BLIP & BLIP & Reverse Learning\\ \cline{3-7}
&  & SDQUR~\cite{xu2024SDQUR} & 2024 & BLIP2 & BLIP2 & Uncertainty Regularization \\ \cline{3-7}
&  & IUDC~\cite{ge2024iudc} & 2024 & CLIP & CLIP & LLM-based Data Augmentation \\ \cline{2-7}

& \multirow{17}{*}{Self-attention-based} & CIRPLANT~\cite{liu2021CIRPLANT} & 2021 & ResNet-152 & - & CIRR Dataset \\ \cline{3-7}
&  & FashionVLP~\cite{goenka2022fashionvlp} & 2022 & ResNet-18, ResNet-50 & BERT & Asymmetric Design \\ \cline{3-7}
&  & FashionViL~\cite{han2022fashionvil} & 2022 & ResNet-50 & BERT & Multi-task Pre-training \\ \cline{3-7}
&  & FaD-VLP~\cite{mirchandani2022fad} & 2022 & CLIP(RN50) & CLIP(RN50) & Multi-task Pre-training \\ \cline{3-7}
&  & AMC~\cite{zhu2023amc} & 2023 & ResNet-50 & LSTM & Dynamic Router Mechanism \\ \cline{3-7}
&  & AACL~\cite{tian2023aacl} & 2023 & Swin Transformer & DistilBERT & Revised Shopping100k Dataset \\ \cline{3-7}
&  & FAME-VIL~\cite{han2023fame} & 2023 & CLIP-B & CLIP-B & Multi-task Pre-training \\ \cline{3-7}
&  & NEUCORE~\cite{zhao2024neucore} & 2023 & ResNet & Bi-GRU & Multi-modal Concept Alignment \\ \cline{3-7}
&  & LMGA~\cite{udhayanan2023lmga} & 2023 & ViT & Visual-BERT & Gradient Attention \\ \cline{3-7}
&  & FashionERN~\cite{chen2024fashionern} & 2024 & CLIP-B & CLIP-B & Modifier Enhancement \\ \cline{3-7}
&  & SDFN~\cite{wu2024sdfn} & 2024 & ResNet-50 & LSTM & Dynamic Router Mechanism \\ \cline{3-7}
&  & SSN~\cite{yang2024ssn} & 2024 & CLIP-B & CLIP-B & - \\ \cline{3-7}
&  & LIMN~\cite{wen2023limn} & 2024 &CLIP-L & CLIP-L & Pseudo Triplet Generation \\ \cline{3-7}
&  & SPRC~\cite{xusentence2024sprc} & 2024 & BLIP2 & BLIP2 & Auxiliary loss \\ \cline{3-7}
&  & VISTA~\cite{zhou2024vista} & 2024 & EVA-CLIP-02-Base & BGE-Base-v1.5 & - \\ \cline{3-7}
&  & SyncMask~\cite{song2024syncmask} & 2024 & VIT-B & BERT & Multi-task Pre-training \\ \cline{3-7}
&  & UniFashion~\cite{zhao2024unifashion} & 2024 & CLIP & CLIP & Multi-task Pre-training \\ \cline{2-7}

& \multirow{2}{*}{Graph-attention-based} & JAMMA~\cite{zhang2020jamma} & 2020 & ResNet-101 & Bi-GRU & - \\ \cline{3-7}
&  & GSCMR~\cite{zhang2021GSCMR} & 2022 & ResNet-101 & Bi-GRU & - \\ \cline{3-7}

\hline

\multicolumn{2}{|c|}{\multirow{2}{*}{Prototype Image Generation-based Fusion}} & SynthTripletGAN~\cite{tautkute2021Synth} & 2021 & ResNet-101 & Bi-GRU & - \\ \cline{3-7}
\multicolumn{2}{|c|}{}& TIS~\cite{zhang2022tis} & 2022 & Inception-v3 & LSTM & - \\ 
 
 \hline

\multicolumn{2}{|c|}{\multirow{7}{*}{Plug-and-play}} & RTIC~\cite{shin2021rtic} & 2021 & ResNet-50 & LSTM & GCNs stream\\  \cline{3-7}
 \multicolumn{2}{|c|}{}& JPM~\cite{yang2021jpm} & 2021 & ResNet-18 & LSTM & Image Difference Alignment \\ \cline{3-7}
 \multicolumn{2}{|c|}{}& GA~\cite{huang2022ga} & 2022 & ResNet-18 & LSTM & Gradient Augmentation \\ \cline{3-7}
 \multicolumn{2}{|c|}{}& VQA4CIR~\cite{feng2023vqa4cir} & 2023 & - & - & Re-rank \\ \cline{3-7}
 \multicolumn{2}{|c|}{}& CIR-MU~\cite{chen2022mu} & 2024 & ResNet-50 & RoBERTa & Uncertainty Modeling \\ \cline{3-7}
 \multicolumn{2}{|c|}{}& CaLa~\cite{jiang2024cala} & 2024 & - & - & Image Difference Alignment \\ \cline{3-7}
  \multicolumn{2}{|c|}{}& SDA~\cite{sda2024} & 2024 & - & - & LLM-based Data Augmentation \\ \cline{3-7}
 \multicolumn{2}{|c|}{}& SPN~\cite{feng2024spn} & 2024 & - & - & Positive Example Generation \\ 
 \hline
\end{tabular}

\label{tab:sup_cir}
}
\end{table*}

\section{Supervised Composed Image Retrieval}
In this section, we first provide the problem statement for the task of supervised CIR, and then
present existing approaches. Generally,  as depicted in Figure~\ref{framework_scir}, existing models involve four key components: feature extraction, image-text fusion, target matching, and data augmentation. The first three are essential components for CIR, while the last one is optional and aimed at enhancing model performance. Existing supervised CIR methods are summarized in Table~\ref{tab:sup_cir}.

\subsection{Problem Statement.} Given a reference image and its modification text, CIR aims to retrieve target images from a collection of gallery images. In the supervised learning setting, existing methods rely on training samples in triplet form, \textit{i.e.}, <reference image, modification text, target image>. Let $\mathcal{D}$ = $\{(I_r,T_m,I_t)_i\}^N_{i=1}$ denote a set of such triples, where $I_r$ is the reference image, $T_m$ is the modification text, $I_t$ signifies the target image, and $N$ is the total number of triplets. Then, based on the training dataset $\mathcal{D}$, existing methods aim to learn a multimodal fusion function that effectively combines the multimodal query $(I_r,T_m)$ and a visual feature embedding function to ensure that the composed query and the corresponding target image are close in the embedding space. This can be formalized as follows,
\begin{equation}
f(I_r,T_m) \rightarrow h(I_t), 
\end{equation}
where $f(\cdot)$ represents the multimodal fusion function mapping the multimodal query to the latent space, while $h(\cdot)$ denotes the feature embedding function for the target image.

\subsection{Feature Extraction}
In the task of CIR, feature extraction plays a crucial role in deriving meaningful embeddings from both the input query and the target image. Since feature extraction has been extensively studied in fields such as natural language processing and computer vision, most existing methods for CIR leverage established textual and visual feature extraction backbones to encode the input query and target image. We categorize these encoder backbones into two primary types: traditional encoders and vision-language pre-trained (VLP) model-based encoders.

\subsubsection{Traditional Encoder.}
For textual feature extraction, commonly used encoders for CIR tasks include RNN-based encoders and transformer-based encoders. Representative RNN-based encoders used in CIR studies are Bidirectional Gated Recurrent Units (BiGRUs)~\cite{cho2014learning} and Long Short-Term Memory networks (LSTMs), which have proven effective in capturing long-term dependencies in text sequences. Specifically, existing CIR studies~\cite{delmas2022artemis,li2023dscn,li2024cmap,li2023manme,tautkute2021Synth,zhang2020jamma, zhang2021GSCMR, wang2024NSFSE,li2023acnet,zhao2024neucore} employ BiGRUs as text encoders to process sequences bidirectionally, enriching feature embedding by capturing context from both past and future tokens. Meanwhile, several studies~\cite{zhang2022eer,huang2023lgli,xu2024alret,zhang2022tis,zhang2023MLCLSAP,yang2023crn,zhang2024mcem,dodds2020maaf,wu2024sdfn, zhang2021mcr} utilize LSTMs, which introduce gated mechanisms to the standard RNN structure, effectively managing long-range dependencies for modification text feature extraction. With the emergence of transformers~\cite{vaswani2017attention}, a growing number of CIR studies~\cite{jandial2022sac,anwaar2021ComposeAE, goenka2022fashionvlp,hu2023provla,han2022fashionvil,tian2023aacl, xu2023ComqueryFormer,song2024syncmask} adopt transformer-based encoders, such as BERT~\cite{devlin2019bert} and its variants (\textit{e.g.}, RoBERTa~\cite{liu2019roberta} and DistilBERT~\cite{sanh2019distilbert}), as their text encoders. These encoders leverage self-attention mechanisms to capture global context across the entire text sequence, enabling parallel processing and producing deeper contextual embeddings. Overall, compared to RNN-based encoders, transformer-based encoders demonstrate superior capabilities for textual embedding in CIR tasks, particularly when pre-trained on extensive corpora.

Similarly, traditional image encoders used in CIR studies can be categorized into CNN-based and transformer-based encoders. CNN-based encoders are initially popular due to their ability to capture spatial hierarchies through convolution operations, preserving crucial spatial information and providing robust hierarchical feature embeddings. Many CIR methods~\cite{vo2019tirg, chen2020val,liu2021mgf,wen2021clvcnet,zhang2023pcasm,delmas2022artemis,zhang2023MLCLSAP,huang2023lgli,xu2024alret,tautkute2021Synth,zhang2024cssnet} extract image features with pre-trained CNN-based encoders, such as ResNet~\cite{krizhevsky2017imagenet}, GoogleNet~\cite{szegedy2015going}, and MobileNet~\cite{howard2017mobilenets}, which yield generalizable feature embeddings by being pre-trained on large-scale datasets like ImageNet~\cite{deng2009imagenet}. 
In contrast to CNN-based encoders, which directly feed the entire image into the encoder, transformer-based encoders redefine image encoding by segmenting images into non-overlapping patches and employing self-attention to model spatial relationships. One commonly used transformer-based encoder for CIR~\cite{song2024syncmask,udhayanan2023lmga} is Vision Transformer (ViT)~\cite{dosovitskiy2021image}, which captures more nuanced visual details with a self-attention mechanism over image patches. Additionally, several CIR methods~\cite{xu2023ComqueryFormer, hu2023provla,tian2023aacl,yang2023crn} employ Swin Transformers~\cite{liu2021swin}, which adopt a window-based self-attention mechanism for local interactions within each window, reducing computational complexity. 
Typically, transformer-based encoders offer superior representational capabilities compared to CNN-based ones, particularly when pre-trained on extensive datasets.

\subsubsection{VLP-based Encoder.}
Recently, advancements in vision-language pre-training have led to the prevalence of VLP-based encoders, which are now preferred for encoding multimodal data due to their ability to align visual and textual modalities. For example, several studies~\cite{chen2024spirit,wen2023tgcir,yang2024ssn,xu2024alret,baldrati2022CLIP4CIR,zhao2022PL4CIR,chen2023ranking,lin2023clip_cd,wen2024dqu,wan2024caff} adopt CLIP~\cite{radford2021learning} as their feature extraction backbone, which leverages contrastive learning on large-scale image-text datasets and demonstrates exceptional flexibility across diverse domains. In addition, some studies~\cite{liu2023rerank,liu2024blip4cir,levy2024case} utilize BLIP~\cite{li2022blip}, which unifies vision-language understanding and generation with a multimodal mixture of encoder-decoder framework.
Moreover, BLIP-2~\cite{li2023blip2} has also been adopted for feature extraction in recent CIR studies~\cite{xusentence2024sprc, xu2024SDQUR}, which bridges the modality gap with a lightweight Querying Transformer (Q-Former) and achieves state-of-the-art performance on various vision-language tasks.
Collectively, these VLP-based encoders provide more robust feature embeddings for CIR tasks, as compared with traditional encoders. 

\subsection{Image-text Fusion}
Once the reference image and modification text in the input query are separately encoded, the subsequent crucial step involves designing an effective image-text fusion strategy to integrate the complementary information from both modalities, in order to precisely represent the input query and conduct the target image retrieval. Towards this end, existing methods can be categorized into three groups: explicit combination-based fusion, neural network-based fusion, and prototype image generation-based fusion.

\subsubsection{Explicit Combination-based Fusion}
The first group of methods aims to accomplish image-text fusion through explicit combination operations, which can be categorized into two types: transformed image-and-residual combination and content-and-style combination.

\textbf{Transformed Image-and-Residual.} The key idea of this group of methods is to keep the image feature as the dominant component, and achieve image-text fusion by learning two parts: the transformed reference image feature and the residual feature.
This group of methods can typically be expressed as \underline{$g\left( \left[\mathit{img};\mathit{txt}\right] \right) \odot \mathit{img} + \mathit{res}$}. Here, $\mathit{img}$ and $\mathit{txt}$ represent the reference image embedding and modification text embedding, respectively. Notably, $\mathit{img}$ and $\mathit{txt}$ may not be the direct output of the feature extraction component. 
To enhance image-text fusion, various types of image/text features, such as global features~\cite{vo2019tirg, chen2020jvsm,chen2023ranking,zhang2024mcem,xu2024alret}, local features~\cite{liu2021mgf,zhang2022eer,lee2021cosmo}, hierarchical features~\cite{chen2020val,kim2021dcnet,wen2021clvcnet,yang2023crn,jandial2022sac,zhang2023MLCLSAP,huang2023lgli,huang2024dwc,chen2024spirit,zhang2024cssnet}, and decoupled features~\cite{wen2023tgcir,chawla2021lsc4cir,zhang2023pcasm} have been explored. The function $g\left(\cdot\right)$ is a neural network that derives the parameters applied to the reference image to achieve the modification operation, and $\odot$ denotes the element-wise multiplication. $\mathit{res}$ refers to the residual offsetting information. 
Based on the method used to obtain $\mathit{res}$, the approaches in this branch can be further divided into two major categories. 
The first category of methods, including TIRG~\cite{vo2019tirg},VAL~\cite{chen2020val}, JVSM~\cite{chen2020jvsm}, MGF~\cite{liu2021mgf}, DCNet~\cite{kim2021dcnet}, CLVC-Net~\cite{wen2021clvcnet}, Css-Net~\cite{zhang2024cssnet}, and CRN~\cite{yang2023crn}, directly fuse the image and text features to derive the residual offsetting information, \textit{i.e.}, \underline{$\mathit{res} = h\left( \left[\mathit{img};\mathit{txt}\right] \right)$}. The most representative method is TIRG~\cite{vo2019tirg}, which designs $g\left(\cdot\right)$ as a gating function with a Sigmoid activation function to adaptively preserve the unchanged information in the reference image, and $h\left(\cdot\right)$ as a simple MLP-based neural network to fuse the image and text features and derive the residual offsetting information. Given its simple yet effective image-text fusion paradigm, the following JVSM, MGF, and DCNet directly adopt the same manner as TIRG does, while VAL, CLVC-Net, and CRN share similar spirits as TIRG, except that they utilize the attention mechanism instead of the gating function. For example, VAL devises a joint-attention mechanism ($g\left(\cdot\right)$) to suppress and highlight the visual content based on the spatial and channel dimensions of the reference image feature maps. Meanwhile, it combines self-attention learning ($h\left(\cdot\right)$) to capture crucial visio-linguistic cues as the residual part. CLVC-Net devises two attention mechanism-based streams to derive $g\left(\cdot\right)$ and $h\left(\cdot\right)$ from both local-wise and global-wise perspectives. Subsequently, these two streams are made to learn from one another with mutual learning in order to obtain a more comprehensive image-text fusion result. In particular, Css-Net adopts two different forms of compositor, namely \underline{$\mathit{img} + \mathit{res}$} and \underline{$\mathit{text} + \mathit{res}$}. The former primarily aims to discern ``what elements to modify'' within the reference image, guided by the modification text; while the latter emphasizes determining ``what to retain'' within the modification text given the reference image. 

Another group of methods adheres to the formula expressed as \underline{$\mathit{res} = h\left( \left[\mathit{img};\mathit{txt}\right] \right) \odot \mathit{txt}$} to learn the residual part. Typically, both $g\left(\cdot\right)$ and $h\left(\cdot\right)$ are scaled to fall within the range of $0$ and $1$. This strategy works on not only preserving the unchanged part within the reference image, but also integrating information from the modification text that is to replace certain aspects. TG-CIR~\cite{wen2023tgcir}, DWC~\cite{huang2024dwc}, AlRet~\cite{xu2024alret}, and EER~\cite{zhang2022eer} all follow this criteria to achieve the image-text fusion. Several methods adopt a most straightforward approach by directly adding $\mathit{img}$ and $\mathit{txt}$, such as MCEM~\cite{zhang2024mcem} and Ranking-aware~\cite{chen2023ranking}. For the sake of convenience, we also classify them into this group by considering $g\left(\cdot\right) = h\left(\cdot\right) = \mathbf{I}$, \textit{i.e.}, identity mapping. 
Likewise, LGLI~\cite{huang2023lgli}, where $g\left(\cdot\right) = \mathbf{I}$ and SAC~\cite{jandial2022sac}, where $h\left(\cdot\right) = \mathbf{I}$) are also categorized into this group. Notably, instead of using element-wise multiplication to transform the image features, SAC relies on the attentional transformation to derive the text-conditioned image representation and then adds it to the modification text representation to obtain the fused feature, which can be expressed as $s\left( \left[\mathit{img};\mathit{txt}\right] \right) + \mathit{txt}$. $s\left(\cdot\right)$ denotes the attentional transformation network. 



\textbf{Content-and-Style.} Considering that ``each image can be well characterized by their content and style''~\cite{chawla2021lsc4cir}, this group of methods~\cite{lee2021cosmo, chawla2021lsc4cir, zhang2023pcasm, chen2024spirit} typically hypothesizes that both the image style and content will be modified in accordance with the modification text. Thereby, modifications are achieved in both the style and content spaces, and they are finally combined to obtain the fusion output. For example, CoSMo~\cite{lee2021cosmo} initially devises a content modulator equipped with a disentangled multimodal non-local block for effecting content modifications. Subsequently, once the content has been adjusted, a style modulator is designed to incorporate the modified style information, ensuring a sequential progression from content to style adaptation.
Differently, LSC4TCIR~\cite{chawla2021lsc4cir} and PCaSM~\cite{zhang2023pcasm} pursue content and style modifications in parallel. Specifically, they decompose the reference image into its corresponding style and content features, then perform semantic replacement within these feature spaces according to the modification text. Finally, the content and style features are fed into the combination module parallel for image-text fusion. 
While SPIRIT~\cite{chen2024spirit} primarily centers on the style aspect. It introduces an explicit definition of style as the commonality and difference among the local patches of an image. With this concept in mind, SPIRIT first segments the reference image into multi-granularity patches. Subsequently, two modules are devised to model the local patches' commonality and difference, respectively. Then the two style features are combined to enrich the representation of the reference image, which is fused with text features to obtain the final query
representation. 

\subsubsection{Neural Network-based Fusion}
In contrast to the methods of the former group, methods in this group rely entirely on neural networks to fuse the reference image and modification text without employing explicit combination operations. These methods can be further categorized into four subgroups: MLP-based, cross-attention-based, self-attention-based, and graph-attention-based approaches.

\textbf{MLP-based.} Methods in this branch~\cite{anwaar2021ComposeAE,baldrati2022Combiner,baldrati2022CLIP4CIR,zhao2022PL4CIR,baldrati2023CLIP4CIR2,li2023dscn,lin2023clip_cd,liu2024blip4cir,wan2024caff,li2023manme,zhang2024cssnet,yan2024shaf,li2024clip,wen2024dqu,wang2024sadn,wang2024NSFSE,delmas2022artemis,li2024cmap,dmot} primarily rely on the multi-layer perceptron (MLP) to fulfill the image-text fusion. For example, PL4CIR~\cite{zhao2022PL4CIR} introduces a multi-stage learning framework to progressively acquire the complex knowledge necessary for multimodal image retrieval and employs an MLP-based query adaptive weighting strategy to dynamically balance the influence of image and text. 
As pioneers in applying CLIP to CIR tasks, Baldrati~\textit{et al.}~\cite{baldrati2022Combiner} introduce a classic multimodal fusion network, \textit{i.e.}, combiner, which integrates CLIP-based reference image and modification text features using MLP-based weighted summing and feature concatenation. Later, they also propose several fine-tuning strategies~\cite{baldrati2022CLIP4CIR, baldrati2023CLIP4CIR2} within this framework to alleviate the domain discrepancy between CLIP's pre-training data and the downstream task data. This combiner network has been directly adopted by many subsequent methods~\cite{liu2024blip4cir, wen2024dqu, wang2024sadn} or refined in later studies~\cite{lin2023clip_cd,wan2024caff}. 
Considering the differing contribution between the input reference image and modification text, DWC~\cite{huang2024dwc} introduces an Editable Modality De-equalizer (EMD). This module employs two modality editors equipped with spatial and word attention mechanisms to refine image and text features, respectively. It then utilizes an MLP-based adaptive weighting module to assign modality weights based on contributions. Additionally, DWC further incorporates a CLIP-based mutual enhancement module, which effectively mitigates modality discrepancies and promotes similarity learning. To achieve a more precise alignment of visual and linguistic features, SHAF~\cite{yan2024shaf} initially employs an attention mechanism for text and image feature realignment at multiple levels when encoding the reference image and modification text using FashionCLIP~\cite{chia2022contrastive}. Subsequently, an MLP-based dynamic feature fusion strategy is used for integrating the multimodal information by emphasizing critical features through weight allocation and feature enhancement mechanisms. 

\textbf{Cross-attention-based.} 
CIR methods~\cite{hosseinzadeh2020lbf, dodds2020maaf, xu2023ComqueryFormer, huang2023lgli, li2023acnet, ge2024iudc, hu2023provla} in this branch primarily adopt cross-attention to model the interaction between each word in the modification text and every local region in the reference image, thereby enhancing their fusion. In this approach, one modality serves as the query, while the other acts as the key and value. Among them, LGLI~\cite{huang2023lgli} introduces a localization mask, derived from the object detection model Faster R-CNN~\cite{RenHG2017}, as an additional input for image-text fusion, enabling precise local modifications over the reference image. 
Recognizing that the generated localization mask may not always be reliable, the authors additionally introduce a channel cross-modal attention mechanism and a spatial cross-modal attention mechanism to effectively localize the to-be-modified regions. 
ACNet~\cite{li2023acnet} proposes a multi-stage compositional framework that sequentially modifies the reference image based on textual semantics. At each stage, the framework first enhances image features using a self-attention layer applied to the image's regional features. It then applies a cross-attention-based relation transformation layer to establish cross-modal associations between image regions and word features.
To address comprehensive modification intent reasoning, IUDC~\cite{ge2024iudc} introduces a dual-channel matching model comprising a semantic matching module and a visual matching module. The semantic matching module utilizes a gate-based attention mechanism to fuse attributes of the reference image, generated by a large language model~(LLM), with the modification text for semantic reasoning. The visual matching module uses affine transformation for image-text fusion, preceded by a cross-attention mechanism to enhance interaction between the two input modalities. These two modules are trained collaboratively, transferring knowledge to each other for mutual enhancement.
In contrast, to capture fine-grained deterministic many-to-many correspondence between the composed query and target, 
SDQUR~\cite{xu2024SDQUR} leverages the Q-former module of BLIP2~\cite{li2023blip2} to achieve adaptive fine-grained image-text fusion between each word in the modification text and every local region in the reference image. Specifically, it incorporates a set of learnable queries that first interact with word tokens of modification text through self-attention layers and then exchange information with visual patch features through cross-attention layers, thereby learning diverse semantic aspects from the input query.
Unlike previous VLP encoder-based methods that compose the multimodal query in a late fusion manner, CASE~\cite{levy2024case} introduces a cross-attention driven shift encoder based on BLIP's image-grounded text encoder, \textit{i.e.}, a BERT encoder with intermediate cross-attention layers. The image is first encoded by ViT and then injected into the cross-attention layers of the shift encoder, enabling early image-text fusion. 

\textbf{Self-attention-based.} 
Contrary to cross-attention, in the self-attention mechanism that has gained prominence with the rise of Transformer, the input sequence simultaneously plays roles of the query, key, and value, for computing the attention scores for each input element. Consequently, self-attention enables the learning of dependencies among different elements within a single sequence. Methods~\cite{mirchandani2022fad,tian2023aacl,han2023fame,zhao2024neucore,udhayanan2023lmga,chen2024fashionern,wen2023limn,song2024syncmask,yang2024ssn,zhao2024unifashion} in this branch typically feed the concatenation of the encoded reference image feature and the modification text feature into a self-attention-based network, like Transformer, to fully learn their interaction for promoting image-text fusion. 
Among them, AACL~\cite{tian2023aacl} refines the standard Transformer encoder with an additive self-attention layer, which utilizes the additive attention mechanism to capture contextual information and selectively suppress or highlight the representation of each token, thereby facilitating the retention and modification of reference image information. 
Different from previous work that treats the modification text as a single description, SSN~\cite{yang2024ssn} treats the modification text as an instruction and hence explicitly decomposes the semantic transformation conveyed by the modification text into two steps: degradation and upgradation. Specifically, it first uses an MLP-based degradation network to degrade the reference image to a visual prototype that only retains to-be-preserved visual attributes. Then, a Transformer-based upgrading network is adopted to upgrade the visual prototype to the final desired target image. Both processes are guided by the modification text. 
Contrary to above methods that compose the multimodal query based on their extracted features, some methods~\cite{liu2021CIRPLANT,goenka2022fashionvlp,xusentence2024sprc,zhou2024vista,han2022fashionvil} first concatenate the encoded reference image features and word tokens from the modification text
and then feed the concatenated token sequence into a transformer-based model for image-text fusion.
Different from the above methods that adopt a single fusion strategy, AMC~\cite{zhu2023amc} and SDFN~\cite{wu2024sdfn} consider multiple fusion strategies, where the self-attention mechanism acts as only one fusion option, and incorporates a dynamic routing mechanism for adaptive image-text fusion. 
Especially, some methods~\cite{mirchandani2022fad,song2024syncmask,han2023fame} focus on addressing various vision-and-language (V+L) tasks in the fashion domain, including cross-modal retrieval and CIR, within a transformer-based foundation model. By pretraining on multiple meticulously designed tasks, their transformer-based frameworks exhibit competitive performance across a range of heterogeneous fashion-related tasks.

\textbf{Graph-attention-based.} Graph attention mechanisms are specifically designed for handling graph-structured data. Unlike the above cross-attention and self-attention, their primary distinction lies in that they deal with data structured as graphs rather than sequences, and they take into account the relationships between edges and nodes within the graph. 
To manipulate the visual features of the reference image according to semantics in the modification text at the attribute level, JAMMA~\cite{zhang2020jamma} leverages the pretrained Faster R-CNN to extract visual attribute features for each reference/target image. These features are then utilized as vertices to construct a graph, where each edge is established based on the relative size and location relationships between two attributes~\cite{darec2019}. Subsequently, JAMMA employs a jumping graph attention network to infuse semantic information from the modification text into the attribute graph, dynamically assigning higher weights to attributes most relevant to the text. Finally, it adopts a global semantic reasoning module, which follows the idea of gate and memory mechanism~\cite{vscitm}, to filter out redundant attributes, thereby yielding a more discriminative global query feature. By jointly modeling the geometric information of the image and the visual-semantic relationship between the input image and text, GSCMR~\cite{zhang2021GSCMR} initially learns cross-modal embedding for the composed query in a geometry-aware way and then rectifies the visual feature under the guidance of the modification text with a multi-head graph attention network~\cite{2018graphattentionnetworks}. 

\subsubsection{Prototype Image Generation-based Fusion}
Apart from the first two groups, some methods aim to achieve multimodal fusion by directly synthesizing a prototype image that satisfies the requirements of the multimodal query. This approach effectively converts CIR into an image-to-image retrieval problem.
Recognizing the lack of interpretability in traditional methods that directly combine inputs into a multimodal query representation for target image retrieval, SynthTripletGAN~\cite{tautkute2021Synth} pioneers integrate the Generative Adversarial Networks (GANs) into CIR, where a triplet loss is used for metric learning. %
In contrast, TIS~\cite{zhang2022tis} introduces a multi-stage GAN-based structure that embeds a retrieval model within a GAN framework. To learn a discriminative composed query feature,
TIS uses two distinct discriminators: one targeting global differences between generated and target images, and the other identifying local modifications in the generated images. 
Unlike the above two methods that concentrate on the CIR task, UniFashion~\cite{zhao2024unifashion} focuses on developing a unified framework leveraging LLMs and diffusion models to enhance the performance of multimodal retrieval and generation tasks with mutual task reinforcement.

\subsection{Target Matching}
The target matching module aims to accurately retrieve images that match the given multimodal query. One fundamental technique for target matching in CIR is metric learning, which establishes a feature space where distances effectively represent semantic similarities or differences. To enhance metric learning, several strategies have been developed, which can be categorized into five groups: basic metric learning, image difference alignment, negative mining, uncertainty modeling, and re-ranking. 

\subsubsection{Basic Metric Learning}
Existing CIR studies primarily utilize three types of loss functions: the batch-based classification (BBC)~\cite{vo2019tirg} loss function, the soft triplet-based loss function, and the hinge-based triplet ranking function.

\textbf{Batch-based classification loss.} This loss function is the most widely used for metric learning in current CIR studies~\cite{vo2019tirg,hosseinzadeh2020lbf,kim2021dcnet,baldrati2022CLIP4CIR,wen2023tgcir,levy2024case,yang2024ssn,jiang2024cala,yan2024shaf,huang2023lgli,xusentence2024sprc,shin2021rtic,zhang2022eer,huang2022ga,li2023acnet,baldrati2023CLIP4CIR2,baldrati2022Combiner}. This loss function aims to bring the query embedding closer to the annotated target image embedding while treating all other target images in the same batch as negative samples, pushing them away from the query embedding.
It is formulated as follows, 
\begin{equation}
L_{BBC} = \frac{1}{B} \sum_{i=1}^{B}\left[ -\log \left(\frac {exp\{\kappa({\phi}^{(i)}, \mathbf{x_t}^{(i)}) /\tau \}}{\sum_{j=1}^{B}exp\{\kappa({\phi}^{(i)} , \mathbf{x_t}^{(j)} ) /\tau \}} \right) \right], 
\label{eq:bbc_loss}
\end{equation}
where the subscript $i$ refers to the $i$-th triplet sample in the mini-batch, ${\phi}$ and $\mathbf{x_t}$ represent the combined feature of the input query and the target feature, respectively. $B$ is the batch size, $\kappa\left( \cdot, \cdot \right)$ serves as the cosine similarity function, and $\tau$ denotes the temperature factor. 

\textbf{Soft triplet-based loss.}
This loss function is a specific variant of the batch-based classification loss function, where a single candidate image from the batch is selected as the negative example at each iteration. It has been widely adopted for metric learning in several CIR studies~\cite{vo2019tirg, hosseinzadeh2020lbf, liu2021mgf, liu2021CIRPLANT, yang2021jpm}. Specifically, this loss function is formulated as follows,
\begin{equation}
L_{ST} = \frac{1}{MB} \sum_{i=1}^{B}  \sum_{m=1}^{M}  \log \{ 1 + exp\{ \kappa({\phi}^{(i)}, \mathbf{x_t}^{(i)}) - \kappa({\phi}^{(i)}, \mathbf{\tilde{x}_{(t,m)}})  \} \}, 
\label{eq:st_loss}
\end{equation}
where the subscript $i$ refers to the $i$-th triplet sample in the mini-batch, $ \mathbf{\tilde{x}_{(t,m)}}$ represents the $m$-th selected negative sample, and $M$ is the repeat times to evaluate every possible set. 

\textbf{Hinge-based triplet ranking loss.} This loss focuses on optimizing hard negative samples and has been adopted by several CIR studies~\cite{zhang2020jamma, zhang2021GSCMR, chen2020jvsm, li2023dscn}. Similar to the above loss, its primary objective is to cluster matched query-target pairs while separating unmatched ones in the embedding space. By concentrating on hard negatives, this loss function this loss function effectively addresses the challenges of high redundancy and slow convergence associated with the random triplet sampling process inherent in soft triplet-based loss functions. Formally, the objective function is defined as follows, 
\begin{equation}
L_{rank} = max[0, \gamma - F({\phi}^{(i)}, \mathbf{x_t}^{(i)}) + F({\phi}^{(i)}, \mathbf{\tilde{x_t}}^{(i)})] 
+ max[0, \gamma - F({\phi}^{(i)}, \mathbf{x_t}^{(i)}) +F({\tilde{\phi}}^{(i)}, \mathbf{x_t}^{(i)})], 
\label{eq:hb_loss}
\end{equation}
where the subscript $i$ refers to the $i$-th triplet sample in the dataset, $F(\cdot)$ denotes the semantic similarity function, $\gamma$ is a margin value, ${\tilde{\phi}}$ and $\tilde{\mathbf{x_t}}$ represent the hard negatives for the positive pair (${\phi}$, $\mathbf{x_t}$). 

Most CIR methods typically rely on single-granularity matching, optimizing the model using one of the three aforementioned loss functions to minimize the distance between the final combined feature and the target image. However, since visual elements vary substantially in scale~\cite{liu2021swin}, several studies~\cite{chen2020val, huang2023lgli, xu2023ComqueryFormer, li2023dscn, li2024cmap, li2023manme} have explored hierarchical matching to improve the alignment between the input query and the target image. These methods start by sampling multi-granular visual features from the visual encoder and separately integrating them with the modification text feature. Then, they minimize the distance between these composed features and the corresponding granularity-level features of the target image based on their respective loss functions to optimize the model. 

\subsubsection{Image Difference Alignment.} 
As mentioned above, mainstream methods model the task as a query-target matching task, \textit{i.e.}, encoding the multimodal query into a single feature and then aligning it with the target image. However, this learning paradigm only explores the most straightforward relationship within each triplet. In fact, beyond this query-target matching relationship, there exists a latent relationship between the reference-target image pair and the modification text. Intuitively, the modification text should capture the visual difference between the reference image and the target image. It acts as an implicit transformation to convert the reference image into the target image. Accordingly, several studies~\cite{kim2021dcnet, zhang2021mcr, yang2021jpm,li2023acnet, jiang2024cala} explore image difference alignment, \textit{i.e.}, aligning the difference between the reference image and target image to the modification text, for boosting the metric learning. To achieve this, several studies~\cite{kim2021dcnet, jiang2024cala, li2023acnet} adapt the conventional BBC loss for image difference alignment as follows, 
\begin{equation}
L_{BBC}^{’} = \frac{1}{B} \sum_{i=1}^{B}\left[ -\log \left(\frac {exp\{\kappa(\mathbf{v_d}^{(i)}, \mathbf{t_m}^{(i)}) /\tau \}}{\sum_{j=1}^{B}exp\{\kappa(\mathbf{v_d}^{(i)} , \mathbf{t_m}^{(j)} ) /\tau \}}  \right) \right], 
\label{eq:imagedifferencebbc_loss}
\end{equation}
where the subscript $i$ refers to the $i$-th triplet sample in the mini-batch, ${\mathbf{v_d}}$ and $\mathbf{\mathbf{t_m}}$ represent the visual difference representation of the reference-target image pair and the modification text representation, respectively. Typically, ${\mathbf{v_d}}$ is derived from neural networks, such as MLP and cross-attention networks.
Differently, JPM~\cite{yang2021jpm} adopts the Mean Squared Error~(MSE) loss to narrow the distance between image differences and text modifications as follows,
\begin{equation}
L_{MSE} = \frac{1}{N} \sum_{i=1}^{N} \parallel \mathbf{v_d}^{(i)} - \mathbf{t_m}^{(i)} \parallel ^2,
\label{eq:mse}
\end{equation}
where the subscript $i$ refers to the $i$-th triplet sample in the dataset.
Different from the above methods, MCR~\cite{zhang2021mcr} formulates image difference alignment as a modification text generation problem. It inputs the features of reference and target images into an LSTM, attempting to generate the modification text directly. The commonly used cross-entropy loss for text generation is adopted to optimize this process. Furthermore, to enhance image difference alignment, NEUCORE~\cite{zhao2024neucore} designs multimodal concept alignment, which targets mining and aligning the visual concepts in the reference and target images with the semantic concepts in modification text. On one hand, NEUCORE extracts keywords from the modification text as semantic concepts, embedding them using GloVe~\cite{pennington2014glove}. On the other hand, it learns the visual concepts present in the reference and target images using a transformer-based model.
Since the modification text is typically concise and contains limited semantic concepts, whereas images convey a wealth of visual concepts, NEUCORE employs an asymmetric loss~\cite{ridnik2021asymmetric} to supervise multimodal concept alignment. This loss function ensures effective alignment by accounting for the inherent imbalance in the richness of concepts between textual and visual modalities, and it is formulated as follows:
\begin{equation}
\left\{
\begin{array}{ll}
\mathbf{s_i} = sigmod(\mathbf{v_{rt}}^{(i)}\cdot \mathbf{w_c}^{(i)} ), \\
L_{asy} = -\frac{1}{N} \big( \sum_{i \in \mathcal{P}}^{}(1-\mathbf{s_i})^{\beta+}  log(\mathbf{s_i}) + \sum_{j\in \mathcal{N}}^{}(\mathbf{s_j})^{\beta-}  log(1-\mathbf{s_j}) \big),
\end{array}
\right.
\label{eq:as_loss}
\end{equation}
where the subscript $i$ refers to the $i$-th triplet sample in the dataset, $\mathbf{v_{rt}}$ and $\mathbf{w_c}$ represent the joint visual concept feature of the reference-target image pair and the embedding of one semantic concept extracted from the modification text, respectively. $\mathcal{P}$ and $\mathcal{N}$ are the positive and negative sets, respectively. ${\beta+}$ and ${\beta-}$ are hyper-parameters that balance the importance of positive and negative concepts, respectively. 

\subsubsection{Negative Mining.}
While the mainstream BBC loss helps models learn associations between composed queries and target images, it treats all other examples within the same batch equally as negative samples. This leads to the issue of false negative samples, as in CIR tasks, a query might correspond to multiple target images, even though only one is annotated as the positive example. 
Moreover, this loss overlooks the varying impact of different negative samples in metric learning. Intuitively, using hard negative samples, \textit{i.e.}, negative examples that are particularly challenging to classify, can significantly benefit model optimization. To address these limitations, researchers have proposed several negative mining techniques.
To deal with the false negative issue, TG-CIR~\cite{wen2023tgcir} first utilizes the visual similarity distribution between the ground-truth target image features and other candidate image features within the batch to regularize the model's metric learning. In addition, NSFSE~\cite{wang2024NSFSE} flexibly learns the boundaries between matched triplets and mismatched triplets using Gaussian distributions, where a flexible threshold is learned to distinguish positive target images from negative ones. 
Furthermore, SADN~\cite{wang2024sadn} first calculates the similarity between the composed query and each candidate target image to select the top-$K$ most relevant candidate images as a neighborhood. It then adaptively aggregates the features of these neighborhood target images to refine the query feature. This incorporation of neighborhood target features effectively mitigates the adverse impact caused by false negative samples. 
To effectively mine hard negative samples, ProVLA~\cite{hu2023provla} introduces a moment queue-based hard negative mining mechanism that uses momentum-based distillation to dynamically store the most recent embeddings of composed images, reference images, and target images. These stored embeddings serve as hard negative samples for triplets, enabling the selection of hard negatives across multiple batches.
Later, instead of constructing conventional query-level hard negative samples, MCEM~\cite{zhang2024mcem} proposes two strategies for generating component-level hard negative samples. The first strategy involves directly replacing the entire modification text in a training triplet to create a hard negative sample. The second strategy narrows the replacement scope by replacing only partial dimensions of the modification text embedding, resulting in more challenging negative samples.
In particular, MCEM introduces a mask vector controlled by a Bernoulli distribution with parameter $p$ for modification text embedding replacement. The parameter $p$ controls the similarity between the newly generated modification text embedding and the original version. 

\subsubsection{Uncertainty Modeling.} 
In the existing CIR datasets, only one target image per query is annotated. However, as mentioned earlier, the inherent ambiguity arising from general modification text often leads to many-to-many relationships between input queries and target images. To address this limitation, Ranking-aware~\cite{chen2023ranking} introduces a novel ranking-aware uncertainty approach, which employs stochastic mappings instead of deterministic ones to capture many-to-many correspondences. Specifically, images and text are encoded not as deterministic features but as distributions in the feature space. This approach optimizes many-to-many ranking through three key components: in-sample uncertainty, cross-sample uncertainty, and distribution regularization.
In contrast, CIR-MU~\cite{chen2022mu} and SDQUR~\cite{xu2024SDQUR} retain the one-to-one matching paradigm while integrating Gaussian-based uncertainty modeling and uncertainty regularization to accommodate more diverse retrieval requirements. Uncertainty modeling simulates the true range of uncertainty within an effective domain, estimated based on the feature distribution within a mini-batch. Additionally, uncertainty regularization prevents the model from excluding potential true positives, thereby improving recall rates.

\subsubsection{Re-ranking.}
As illustrated in Figure~\ref{framework_scir}, most existing CIR methods primarily adopt dual branches: one branch encodes the query, while the other encodes the target image to perform target image retrieval. This architecture ensures efficient inference since the embeddings of all candidate target images can be pre-computed. The model then only needs to embed the given test query and compare it with the pre-computed candidate image embeddings.
However, this approach relies solely on the metric learning module with the BBC loss function to regulate the target image and the query. Due to issues like false negatives and modification ambiguity, the retrieval performance of current CIR models still has room for improvement. To address this, several studies have proposed re-ranking techniques for CIR tasks.
For instance, Re-ranking~\cite{liu2023rerank} introduces a dual-encoder architecture to re-rank the initial retrieval results obtained by conventional dual-branch CIR models. Specifically, one encoder jointly encodes the given query and each candidate target image, while the other jointly encodes the modification text and each candidate target image. An MLP network then fuses the outputs of the two encoders to compute the final ranking score. A contrastive loss, similar to the BBC loss, is used to optimize the re-ranking module. This strategy allows each candidate's target image to interact with the given query more deeply and comprehensively. Importantly, since re-ranking is performed on a subset of candidate images selected by conventional dual-branch CIR models, this sophisticated strategy is computationally feasible during inference.
Additionally, VQA4CIR~\cite{feng2023vqa4cir} re-ranks retrieval results by querying a multimodal large language model (MLLM), \textit{e.g.}, LLaVA~\cite{liu2024visual}, to determine whether the candidate images contain the desired attributes specified by the modification text. This method serves as a post-processing approach that can be seamlessly integrated into any existing CIR model.

\subsection{Data Augmentation}
As mentioned earlier, existing CIR datasets typically consist of triplets in the form of \textless \emph{reference image, modification text, target image}\textgreater. However, creating such training samples is both expensive and labor-intensive, which significantly limits the size of benchmark datasets. As a result, previous research relying solely on these limited samples has faced overfitting issues to some extent and demonstrated poor generalization capabilities.
To overcome this challenge, researchers have proposed various data augmentation strategies.
\begin{itemize}
    \item \textbf{Image replacement-based}. CLIP-CD~\cite{lin2023clip_cd} introduces a CLIP visual similarity-based data augmentation method that replaces the reference or target images in triplets with visually similar alternatives to generate pseudo triplets, effectively enlarging the dataset. Notably, it establishes lower and upper similarity thresholds to ensure the quality and relevance of the augmented samples. 
    \item \textbf{IDC model-based}. Many potential reference-target image pairs in existing datasets remain unlabeled, despite being highly similar and differing only in minor properties. To utilize these unlabeled pairs and enhance model performance, LIMN+~\cite{wen2023limn} introduces an iterative dual self-training paradigm. This approach employs the dual model of CIR—specifically, an image difference captioning (IDC) model~\cite{h2018learning}—to automatically annotate these pairs, generating pseudo-triplets for improved model training. To ensure the quality of the pseudo-triplets generated in each iteration, the CIR model trained in the previous iteration filters out triplets with low query-target matching scores.
    \item  \textbf{LLM-based}. Leveraging the advanced image comprehension capabilities of the large multi-modal model GPT4V, IUDC~\cite{ge2024iudc} employs specially designed prompts to guide GPT4V in generating attribute-level labels for each image in the training triplets. These attribute-level labels are combined with TF-IDF~\cite{salton1988term} features to identify potential reference-target image pairs. Subsequently, ChatGPT is used to generate corresponding modification texts based on the attribute-level labels of these image pairs, enabling the construction of a large volume of triplet data. Additionally, SDA~\cite{sda2024} uses ChatGPT to generate pseudo modification texts by editing specific attributes of the original modification text. GPT4V then generates a target image based on the original reference image and the generated modification text, thereby creating new triplets to assist model training. 
    \item \textbf{Query unification-based}. To fully utilize VLP models like CLIP for mitigating overfitting, DQU-CIR~\cite{wen2024dqu} introduces two raw query unification methods: text-oriented query unification and vision-oriented query unification. In text-oriented query unification, the modification text is combined with the textual description of the reference image, extracted using the VLP model BLIP-2, to create a purely textual query. In vision-oriented query unification, key modification words are directly written onto the reference image at the pixel level, forming a purely visual query. Notably, the image encoder of the VLP model demonstrates strong Optical Character Recognition (OCR) capabilities, enabling it to effectively process such modified images.
    \item \textbf{Reverse objective-based}. CASE~\cite{levy2024case} and BLIP4CIR~\cite{liu2024blip4cir} expand the dataset by incorporating a reverse retrieval objective, \textit{i.e.}, retrieving the reference image given the modification text and the target image. They achieve this by introducing a reverse objective token, such as ``[REV]'' and ``[Backward]'', to specify the retrieval direction. This reverse objective intuitively encourages the model to learn the shift vector between the reference image and the target image from both directions, enhancing its ability to generalize across tasks.
    \item  \textbf{Gradient-based}. Unlike the aforementioned data augmentation methods, GA~\cite{huang2022ga} improves the model's generalization ability through gradient augmentation rather than raw data augmentation. Specifically, GA comprises two components: explicit adversarial gradient augmentation and implicit isotropic gradient augmentation. Explicit adversarial gradient augmentation introduces a gradient-oriented regularization term into the loss function to simulate adversarial sample training. Implicit isotropic gradient augmentation increases triplet diversity by modifying gradients according to the principle of isotropy.
\end{itemize}

\begin{table*}
    \scriptsize
    \centering
    \caption{\textbf{Summarization of main zero-shot composed image retrieval approaches.}}
  \label{tab: zs_CIR}
    \begin{tabular}{|c|c|c|c|c|}
    \hline
      Category & Method & Year & Encoder & Key Aspect\\
    \hline 
    
    \multirow{8}{*}{Textual-inversion-based} 
    & Pic2Word~\cite{pic2word} & 2023  & CLIP-L& Coarse-grained Inversion \\ 
    \cline{2-5}
    & SEARLE~\cite{searle} & 2023  & CLIP-B/L & Coarse-grained Inversion \\ 
    \cline{2-5}
    & iSEARLE~\cite{isearle} & 2024  & CLIP-B/L & Coarse-grained Inversion \\
    \cline{2-5}
    & KEDs~\cite{keds} & 2024  & CLIP-L & Knowledge Enhancement\\
    \cline{2-5}
    & Context-I2W~\cite{context_i2w} & 2024  & CLIP-L & Context-dependent Inversion\\
    \cline{2-5}
    & FTI4CIR~\cite{fti4cir} & 2024  & CLIP-L & Fine-grained Inversion \\
    \cline{2-5}
    & ISA~\cite{isa} & 2024  & BLIP & Adaptive Inversion\\
    \cline{2-5}
    & LinCIR~\cite{lincir} & 2024  & CLIP-L/H/G & Self-masking Projection\\
    
    \hline
    
    \multirow{9}{*}{Pseudo-triplet-based} 
    & TransAgg~\cite{transagg} & 2023  & BLIP-B, CLIP-B/L & Transformer-based\\
    \cline{2-5}
    & HyCIR~\cite{hycir} & 2024  & BLIP-B, CLIP-B & Additional Training Stream \\
    \cline{2-5}
    & MCL~\cite{mcl} & 2024  & CLIP-L & MLLM-based\\
    \cline{2-5}
    & MagicLens~\cite{zhang2024magiclens} & 2024  & CoCa-B/L, CLIP-B/L & Transformer-based\\
    \cline{2-5}
    
    & RTD~\cite{rtd} & 2024  & -& Target-Anchored Contrastive Learning\\
    \cline{2-5}
    & CompoDiff~\cite{compodiff} & 2024  & CLIP-L/G & Diffusion-based\\
    \cline{2-5}
    & PVLF~\cite{pvlf} & 2024  & BLIP & V\&L Prompt learning\\
    \cline{2-5}
    
    & MTI~\cite{mti} & 2023  & CLIP-B/L & Masked Learning\\
    \cline{2-5}
    & PM~\cite{pm} & 2024  & CLIP-L & Masked Learning\\

    \hline
    
    \multirow{5}{*}{Training-free} 
    & CIReVL~\cite{cirevl} & 2023  & CLIP-B/L/G & Language-level Reasoning\\
    \cline{2-5}
    & GRB~\cite{grb} & 2023  & BLIP2 & Coarse-Fine Reranking\\  
    \cline{2-5}
    & LDRE~\cite{ldre} & 2024  & CLIP-B/L/G & Divergent Reasoning\\ 
    \cline{2-5}
    & SEIZE~\cite{seize} & 2024  & CLIP-B/L/G & Divergent Reasoning\\
    \cline{2-5}
    & Slerp~\cite{slerp} & 2024  & BLIP-L, CLIP-B/L & Spherical Linear Interpolation\\
    \cline{2-5}
    & WeiMoCIR~\cite{weimocir}& 2024  & CLIP-L/H/G & Weighted Modality Fusion\\ 
    \hline
    \end{tabular}
\end{table*}

\section{Zero-shot Composed Image Retrieval}
Although supervised CIR approaches have achieved favorable performance, they rely heavily on annotated triplets in the form of \textless \emph{reference image, modification text, target image}\textgreater \hspace{0.2em}for training. However, annotating modification text for each possible \textless \emph{reference image, target image}\textgreater \hspace{0.2em}pair is a time-consuming process. To reduce the dependence on labeled datasets, Pic2Word~\cite{pic2word} introduces ZS-CIR, which aims to perform retrieval without requiring any annotated training triplets. 
Existing ZS-CIR methods can be broadly categorized into three groups: textual-inversion-based, pseudo-triplet-based, and training-free. 
For ease of reference, Table~\ref{tab: zs_CIR} summarizes these methods across the three categories.

\subsection{Textual-inversion-based}
Methods~\cite{pic2word,searle,isearle,context_i2w,fti4cir,isa,keds,lincir} in this category begin by using text inversion technology~\cite{cohen2022my, gal2023an} to map reference image embeddings into textual token representations. These tokens are then combined with those from the modification text to form a unified query. The query is subsequently encoded using the text encoder of a VLP model, like CLIP, enabling image-text fusion.

 \subsubsection{Coarse-grained textual inversion.}
 Pic2Word~\cite{pic2word} is a pioneering method in this category, introducing the task of ZS-CIR. It leverages a collection of unlabeled images to train a lightweight mapping network that transforms image embeddings, obtained from the CLIP visual encoder, into token embeddings compatible with the CLIP text encoder. In this way, each image can be represented by a pseudo-token-based sentence, such as ``a photo of $S^*$'', where $S^*$ denotes a learnable pseudo-word. The mapping network is optimized using a contrastive loss between the image feature and its corresponding pseudo-token-based sentence embedding.
 Around the same time, SEARLE~\cite{searle} proposed two approaches: an optimization-based textual inversion method (OTI) and a mapping network-based method, both aimed at learning pseudo-word tokens that encapsulate the visual content of each image. Unlike Pic2Word, these methods incorporate category-based semantic regularization to align pseudo-word tokens with the CLIP token embedding space, ensuring compatibility with real textual tokens.
 Building on SEARLE, iSEARLE~\cite{isearle} introduces Gaussian noise to text features during OTI to reduce the modality gap between text and image. Moreover, to enhance the mapping network's ability to capture visual contents, iSEARLE implements a similarity clustering-based hard negative sampling strategy, ensuring that each training batch contains a proportion of visually similar images.
 Furthermore, KEDs~\cite{keds} introduces a Bi-modality Knowledge-guided Projection network (BKP), which leverages an external database to provide relevant image-caption pairs as knowledge, enriching the mapping function and improving its generalization ability. Additionally, similar to SEARLE, recognizing the challenges of aligning pseudo-word tokens with real text concepts using only image contrastive training, KEDs introduces an additional training stream that explicitly aligns pseudo-word tokens with semantics using pseudo triplets uncovered from image-caption pairs. Overall, the aforementioned methods perform coarse-grained textual inversion, which simply converts the input image into a single general pseudo-word token, utilizing the entire visual content of the image without differentiation. 

\subsubsection{Fine-grained textual inversion.}
Beyond the above methods, some studies~\cite{context_i2w, keds, fti4cir, isa} have explored fine-grained textual inversion to enhance the performance of ZS-CIR.
For instance, instead of converting the entire visual content of an image, Context-I2W~\cite{context_i2w} adaptively selects caption-relevant content for textual inversion using a context-dependent word mapping network. It employs an intent view selector to map the given image to a task-specific manipulation view and a visual target extractor to collect content related to the specific view.
Instead of converting an image into a single pseudo-word token, FTI4CIR~\cite{fti4cir} maps the image into a subject-oriented pseudo-word token along with several attribute-oriented pseudo-word tokens to comprehensively represent the image in textual form. Additionally, it introduces a tri-wise semantic regularization method based on BLIP-generated image captions, aligning the fine-grained pseudo-word tokens with the real-word token embedding space.
ISA~\cite{isa}, akin to FTI4CIR, converts an image into a series of sentence tokens rather than a single pseudo-word token. It incorporates a spatial attention mechanism-based adaptive token learner to select prominent visual patterns from the image. Moreover, ISA adopts an asymmetric architecture to optimize deployment in resource-constrained environments.
While these methods exhibit promising generalization capabilities to unseen datasets, they rely on a fixed pre-defined text prompt (\textit{e.g.}, ``a photo of $S^*$'') during training. This limitation reduces their ability to handle diverse textual conditions encountered in real-world applications. To address this issue, LinCIR~\cite{lincir} introduces a self-masking projection, which trains a language-only mapping network capable of projecting a given text into a pseudo-token-based embedding by flexibly replacing ``keywords'' (\textit{i.e.}, consecutive adjectives and nouns) in the text with the projected latent embedding of the text. By minimizing the MSE loss between the pseudo-token-based embedding and the projected latent text embedding, the pseudo-tokens of ``keywords'' effectively encapsulate the essential information of the input text. To further bridge the modality gap during inference, LinCIR introduces a random noise addition strategy, adapting the language-only mapping network to handle visual input seamlessly. 

\subsection{Pseudo-triplet-based}
Studies in this category aim to address the ZS-CIR with automatic pseudo-triplet generation methods. 
Existing pseudo-triplet generation methods for ZS-CIR can be categorized into two main classes: LLM-based triplet generation and mask-based triplet generation. 

\subsubsection{LLM-based Triplet Generation.} 
To reduce reliance on manual annotation, several studies~\cite{transagg, hycir, zhang2024magiclens, compodiff, mcl, rtd, pvlf} leverage the advanced logical reasoning capabilities of  LLMs to automatically generate pseudo triplet data. 

\begin{itemize}
    \item \textbf{Image-text-based}. Among these methods, TransAgg~\cite{transagg} is the first to use LLMs for generating pseudo-triplets from a set of image-caption pairs, such as those in the Laion-COCO dataset\footnote{https://laion.ai/blog/laion-coco/}. Specifically, given an image-text pair, it treats the image as the reference image and generates both the modification text and target image caption based on the provided caption, using either a carefully crafted template or an LLM. Drawing inspiration from~\cite{liu2021image}, TransAgg introduces eight types of semantic operations: cardinality, addition, negation, direct addressing, compare\&change, comparative statement, conjunction-based statements, and viewpoint, which guide the modification text generation. It then uses the generated target image caption as a query to retrieve relevant images from the Laion-COCO dataset by calculating the semantic similarity between the target caption and each image's caption. These retrieved images serve as the target images for forming pseudo-triplets, alongside the reference image and generated modification text.
MCL~\cite{mcl} follows a similar pseudo-triplet generation strategy to TransAgg~\cite{transagg}, but instead of using the target image retrieved based on the generated target caption, it directly uses the CLIP feature of the generated target caption as supervision for training the CIR model.
   \item \textbf{Image-based}. Instead of relying on image-text pairs, HyCIR~\cite{hycir} generates pseudo-triplets purely from an unlabeled image dataset, such as COCO~\cite{lin2014microsoft}. The process involves four steps: 1) extracting potential reference-target image pairs; 2) using an image captioning model to generate captions for both images; 3) generating modification text using an LLM based on the two captions; and 4) filtering out triplets with low semantic similarity. Notably, to ensure the generated triplet data is compatible with previous mainstream ZS-CIR methods, such as the textual-inversion-based methods, HyCIR extends the existing Pic2Word method. It introduces an additional training stream that integrates the pseudo-word tokens mapped from the reference image with the modification text, forming a unified text query. This query is then used to supervise both the text query representation and the target image representation using a contrastive loss.
Instead of relying on visual similarity, MagicLens~\cite{zhang2024magiclens} extracts potential reference-target image pairs by mining images from the same webpage, where implicit relationships often exist. It annotates each image with detailed descriptions, including Alt-texts\footnote{https://en.wikipedia.org/wiki/Alt_attribute}, image content annotation (ICA) labels\footnote{https://cloud.google.com/vision/docs/labels}, and captions generated by a large multimodal model, PaLI~\cite{chen2023pali}. Finally, PaLM2~\cite{anil2023palm} is used to generate open-ended modification text for each image pair based on their detailed descriptions. To ensure the generated modification text is logical, techniques such as instruction-following~\cite{chung2024scaling}, few-shot demonstrations~\cite{brown2020language}, and chain-of-thought prompting~\cite{wei2022chain} are employed.
\item \textbf{Text-based}. Inspired by the powerful generation capabilities of diffusion models, Gu et al.~\cite{compodiff} propose a generative approach to construct pseudo triplets. Specifically, they aim to first generate text triplets, structured as~\textless \emph{reference caption, modification text, target caption}\textgreater using two strategies: 1) collecting a large number of captions from existing caption datasets and generating the target caption by substituting keywords in the reference caption, while deriving the modification text based on a randomly sampled pre-defined template; 2) generating the text triplets with an LLM (\textit{i.e.}, OPT-6.7B~\cite{zhang2022opt}), which is fine-tuned with text triplets from existing image editing study InstructPix2Pix~\cite{brooks2023instructpix2pix}. Subsequently, the reference and target images are generated based on their respective captions using a text-to-image generation model, \textit{e.g.}, StableDiffusion~\cite{rombach2022high}. It is worth noting that this work also develops a diffusion-based CIR method, CompoDiff, which can handle various modification cases and enable control over modification strength. Additionally, the textual triplets generated by this work are adopted by a subsequent work~\cite{rtd}, which designs a plug-and-play target-anchored text contrastive learning method for finetuning the CLIP text encoder, thereby improving the performance of textual-inversion-based ZS-CIR.
\end{itemize}

\subsubsection{Mask-based Triplet Generation.}
While the aforementioned LLM-based triplet construction methods efficiently generate large volumes of pseudo-triplet data, they often require excessive computational resources. To address this issue, some researchers have explored mask-based triplet generation strategies~\cite{mti, pm}, which are more resource-efficient. MTI~\cite{mti} is a representative work in this area. Specifically, given an image-caption pair, MTI treats the given image as the target image and randomly masks certain portions of it to derive the corresponding reference image. The provided caption, which encapsulates the predominant content of the image, is then used as the modification text to help reconstruct the masked image back to its original form. In contrast to random masking, PM~\cite{pm} introduces a Class Activation Map (CAM)-guided masking strategy~\cite{chefer2021generic} to better mimic the complementary roles of the reference image and modification text in CIR. Specifically, PM begins by replacing the first noun in the given caption with ``[REMOVE]'' to create the modification text. It then calculates the CAM matrix for the given image, identifying the regions most relevant to the masked noun. These regions are then masked. Unlike MTI, which uses simple color blocks to mask the image, PM replaces the masked regions with corresponding regions from another image within the same batch, ensuring the completeness of the reference image.

\subsection{Training-free}
Given their flexibility and scalability, an increasing body of research is focused on solving ZS-CIR in a training-free manner through modular combinations. These approaches leverage existing models, such as LLMs and VLP models, to address the CIR task without the need for additional model training. Existing methods can be broadly categorized into two branches: one focuses on transforming the CIR task into a caption-to-image retrieval task using powerful LLMs, which can then be handled by the pre-trained encoders of VLP models. The other branch focuses on directly mining the pre-trained common embedding space of VLP models to address the CIR task.

\subsubsection{Task Transformation.} 
Methods in this branch employ LLMs to generate target image captions, which are then used to retrieve the target image via the pre-trained encoders of VLP models. For example, CIReVL~\cite{cirevl} introduces a modular framework where a VLP model generates a description of each reference image. Subsequently, an LLM combines these descriptions with the modification text to infer the target image's caption, which serves as the basis for image retrieval.
Due to the nature of CIR, where the reference image and modification text often involve conflicting semantics, the inferred target image caption may contain concepts that should not appear in the target image. For instance,  given the modification text ``change the dog to a cat", the concept of ``dog" should not appear in the target image.  To address this, GRB~\cite{grb} enhances CIReVL by introducing a Local Concept Re-ranking (LCR) mechanism to ensure the retrieved images contain the correct local concepts. Specifically, GRB performs an initial retrieval based on the inferred target caption, then extracts local concepts that should be present in the target image using an LLM (\textit{e.g.}, ChatGPT4-turbo) from the modification text. To verify the presence of these local concepts, GRB uses LLaVA for Visual Question Answering (VQA) on the top-$K$ retrieved results. The predicted probabilities for outputting text ``yes'' and ``no''  are then used as local scores to re-rank the retrieved images. Moreover, since CIR is inherently a fuzzy retrieval task—where the target image's semantics are not fully captured by the input query—LDRE~\cite{ldre} introduces an LLM-based divergent compositional reasoning approach. This method generates multiple diverse target captions instead of a single one, capturing a wider range of possible semantics within the target image. To complete the ZS-CIR task, LDRE also introduces a divergent caption ensemble technique to combine the CLIP embeddings of these generated captions and retrieve the target image accordingly.

\subsubsection{Pre-trained Space Mining.} 
This branch of training-free methods focuses on leveraging the pre-trained common embedding space of VLP models, such as CLIP and BLIP. These VLP models are primarily optimized using a normalized temperature-scaled cross-entropy loss with cosine similarity, which results in the image and text embeddings residing on a joint hypersphere with the radius determined by the scaling factor (\textit{i.e.}, the temperature parameter). 
WeiMoCIR~\cite{weimocir} directly employs a simple weighted sum to combine the reference image feature, extracted from the visual encoder, and the modification text feature, extracted from the text encoder, to derive the query feature. To improve retrieval performance, WeiMoCIR calculates the score of each candidate image by considering both query-to-image and query-to-caption similarities. It utilizes an MLLM, such as Gemini~\cite{team2023gemini}, to generate multiple captions for each candidate image, providing different perspectives of the image.
In contrast, Slerp~\cite{slerp} applies spherical linear interpolation~\cite{shoemake1985animating} to obtain a fused embedding by calculating intermediate embeddings of the reference image embedding, denoted as $v$, and modification text embedding, denoted as $t$, both derived by VLP encoders. It is formulated as follows: 
\begin{equation}
Slerp(v,t;\alpha) = \frac{\sin((1 - \alpha)\theta)}{\sin(\theta)} \cdot v + \frac{\sin( \alpha)\theta)}{\sin(\theta)} \cdot t,
\end{equation}
where $\theta = \cos^{-1}(v \cdot w)$, and $\alpha$ is a balancing scalar value within the range of [$0$, $1$]. The fused embedding obtained through spherical linear interpolation can be directly used for target image retrieval. 
\label{sec: meth_o}

\begin{table*}
  \scriptsize
  \centering
  \caption{\textbf{Summary of representative approaches for related tasks on composed image retrieval.}}
  \label{tab: task_based_ir}
    \begin{tabular}{|c|c|c|c|c|c|}
    \hline
      Related Task & Method & Year & Visual Encoder & Text Encoder & Key Aspect\\
    \hline 
    
    \multirow{7}{*}{Attribute-based} 
    & AMNet~\cite{amnet} & 2017 & Alex, VGGNet & - & Memory-Augmented \\ 
    \cline{2-6}
    & EItree~\cite{eitree} & 2018 & ResNet-50 & BLSTM & EI-tree \\ 
    \cline{2-6}
    & FashionSearchNet~\cite{FashionSearchNet} & 2018 & AlexNet & - & Attribute Localization \\
    \cline{2-6}
    & EMASL~\cite{EMASL} & 2018 & AlexNet & - & Attribute Localization \\
    \cline{2-6}
    & AMGAN~\cite{amgan} & 2020 & Generator-Encoder & - & GAN-based \\
    \cline{2-6}
    & ADDE~\cite{adde} & 2021 & AlexNet, ResNet 18 & - & Attribute-driven Disentangled \\
    \cline{2-6}
    & FIRAM~\cite{firam} & 2021 & StyleGAN & - & GAN-based \\  

    \hline
    
    \multirow{4}{*}{Sketch-based}
   & TSFGIR~\cite{tsfgir} & 2022 & CNN  & LSTM & Quadruplet Deep Network\\
    \cline{2-6}
    & TASK-former~\cite{taskformer} & 2022 & CLIP-B & CLIP-B & Auxiliary Tasks Learning\\
    \cline{2-6}
    & SceneTrilogy~\cite{SceneTrilogy} & 2023 & VGG-16  & Bi-GRU & Conditional Invertible Networks \\
    \cline{2-6}
    & STNET~\cite{stnet} & 2024 & CLIP & CLIP & Auxiliary Tasks Learning\\
    \cline{2-6}
    & STD4FGIR~\cite{STD4FGIR} & 2024 & CLIP-L & CLIP-L & Textual-inversion-based \\
    
    \hline

    \multirow{2}{*}{Remote Sensing-based} 
    & WEICOM~\cite{weicom} & 2024 & CLIP-L, RemoteCLIP-L & CLIP-L, RemoteCLIP-L & Weighted Average\\
    \cline{2-6}
    & SHF~\cite{shf} & 2024 & ResNet-50 & LSTM & Hierarchical Fusion\\   

    \hline

    \multirow{5}{*}{Dialog-based} 
    & DIIR~\cite{guo2018dialog} & 2018 & ResNet-101 & MLP+CNN & Reinforcement Learning\\
     \cline{2-6}
    & CFIR~\cite{cfir2021} & 2021 & ResNet-101, ResNet-152 & self-attention blocks & Mutual Attention Strategy \\
    \cline{2-6}
    & FashionNTM~\cite{fashionntm} & 2023 & FashionVLP & FashionVLP & Cascaded Memory\\  
    \cline{2-6}
    & IRR~\cite{irr} & 2023 & FashionBERT& CLIP-B & Iterative Sequence Refinement\\ 
    \cline{2-6}
     & Fashion-GPT~\cite{fashiongpt} & 2023 & Swin Transformer& RoBERTa& LLM-integration\\
     \cline{2-6}
    & LLM4MS~\cite{llm4ms} & 2024 & CLIP-L & Flan T5 XL & LLM-integration\\
    
    \hline
    
    \multirow{4}{*}{Video-based} 
    & CoVR-BLIP~\cite{Covr} & 2024 & CLIP, BLIP & CLIP, BLIP & BLIP-based Fusion\\
    \cline{2-6}
    & ECDE~\cite{ecde} & 2024 & BLIP & BLIP & Video Contextual Complement  \\
    \cline{2-6}
    & TFR-CVR~\cite{tfrcvr} & 2024 & BLIP, CLIP-L & BLIP, CLIP-L & Language-level Reasoning \\
    \cline{2-6}
    & CoVR-BLIP2~\cite{CoVR2} & 2024 & BLIP-2 & BLIP-2 & BLIP2-based Fusion\\ 
    
    \hline
    \end{tabular}
\end{table*}

\section{Related Tasks of Composed Image Retrieval}
Beyond the primary task of CIR, researchers have explored various related tasks that cater to diverse retrieval needs in real-world scenarios. Here, we present five representative related tasks that involve different types of multimodal queries: attribute-based, sketch-based, remote sensing-based, dialog-based, and video-based. For ease of reference, we summarize the methods proposed for  these tasks in Table~\ref{tab: task_based_ir}. 

\subsection{Attribute-based}
Before the formal introduction of natural language modification-based CIR, another flexible image retrieval task—where the query involves a reference image alongside attribute-based modifications—had garnered significant attention. This task, constrained to predefined attributes, primarily finds applications in domains with structured attribute sets, such as fashion and face image retrieval.
Zhao~\textit{et al.}~\cite{amnet} pioneered this task by proposing AMNet, which comprises a memory block for storing the representations of various attribute values. The attribute manipulation is achieved by directly retrieving specific attribute representations from the memory block and then fusing them with the representation of the reference image to retrieve target images. Similarly, ADDE~\cite{adde} also incorporates a memory block for storing the attribute values. Unlike AMNet, ADDE devises an attribute-driven disentanglement module, which uses attributes as supervised signals to guide the learning of disentangled image representations. Through indexing the appropriate attribute representations from the memory block, the disentangled representations can be modified by removing, retaining, or adding specific attribute values, enabling effective attribute manipulation. Recognizing that different attributes are correlated with different regions in fashion images, FashionSearchNet~\cite{FashionSearchNet} employs attribute activation maps to extract region-specific attribute features. Each fashion item can be represented by a set of attribute features, and the attribute manipulation is accomplished by directly replacing the specific attribute feature. Sharing a similar spirit, EMASL~\cite{EMASL} extracts different part features for the acquisition of attribute features based on the designed rules. For example, the upper part is associated with the collar, and the side parts are correlated with the left/right sleeves. 
EITree~\cite{eitree} enhances interpretability by organizing fashion concepts into hierarchical tree structures. Guided by this tree structure, EITree generates meaningful image representations where each dimension corresponds to a specific fashion concept. This structure allows for seamless integration of concept-level user feedback, such as attribute manipulation, into the interpretable representation of fashion items.
Building on the success of generative models in image editing, AMGAN~\cite{amgan} introduces an end-to-end generative attribute manipulation framework. This framework generates a prototype image that aligns with user-desired attribute modifications on a reference image to improve target image retrieval. AMGAN consists of a generator and a discriminator: the generator employs visual-semantic and pixel-wise consistency constraints, while the discriminator incorporates semantic learning for precise attribute manipulation and adversarial metric learning to enhance fashion search effectiveness.
Focusing on face image retrieval, FIRAM~\cite{firam} leverages GANs in its framework design. Unlike AMGAN, which generates a prototype image directly, FIRAM aims to learn sparse and orthogonal basis vectors within StyleGAN's latent space. This approach disentangles attribute semantics, allowing for independent attribute adjustment and preference assignment.

\subsection{Sketch-based} 
Many studies have focused on combining sketch and text descriptions to assist users in retrieving target images. Among these, TSFGIR~\cite{tsfgir} pioneers the exploration of complementarity between text and sketch modalities. It introduces a multi-modal quadruplet deep neural network that encourages both the input sketch and text to be closer to the corresponding positive target image than to any negative target image. Additionally, this work contributes a dataset consisting of $1,374$ sketch-photo-text triplets for shoes.
To model more explicit interactions between sketch and text inputs, TASK-former~\cite{taskformer} extends the dual-encoder-based CLIP framework by incorporating a sketch encoder and two new pretraining objectives: multi-label classification, enabling the three encoders to recognize objects, and caption generation, which reconstructs the input text from joint sketch-text embeddings (computed as the simple addition of the text and sketch embeddings). For more complex scenarios involving rough sketches paired with complementary text descriptions, Gatti \textit{et al.}~\cite{stnet} present a dataset comprising approximately 2M queries and 108K natural scene images. They also propose a multimodal transformer-based framework, STNET. Like TASK-former, STNET employs three CLIP encoders for text, sketch, and image modalities. However, beyond contrastive learning, it incorporates three task-specific pretraining objectives: (a) object classification, (b) sketch-guided object detection, and (c) sketch reconstruction.
Distinct from the aforementioned methods, STD4FGIR~\cite{STD4FGIR} draws inspiration from textual inversion techniques by transforming the input sketch into a pseudo-word token, allowing the multimodal input to be directly encoded by the CLIP text encoder. To address the challenge of costly data annotation for sketch-based fine-grained image retrieval, STD4FGIR is trained using only sketch-image pairs. It treats the difference between the image embedding and the sketch embedding as a proxy for the missing text query and introduces a compositionality constraint to model this relationship. To achieve fine-grained matching between the composed query and the target image, STD4FGIR incorporates two innovative loss functions: a region-aware triplet loss for precise alignment and a sketch-to-photo reconstruction loss to enhance representation learning.
It is worth noting that STD4FGIR can be applied to various applications, such as object-sketch-based scene retrieval, domain attribute transfer, and sketch+text-based fine-grained image generation. In parallel, SceneTrilogy~\cite{SceneTrilogy} focuses on learning a flexible joint embedding capable of supporting any combination of modalities—sketches, images, or text—as queries for diverse retrieval and captioning tasks. Leveraging conditional invertible networks, SceneTrilogy disentangles the feature representation of the input sketch, text, or photo into two components: a modality-agnostic part and a modality-specific part. The modality-agnostic parts across all three modalities are aligned using contrastive loss to enable cross-modal retrieval, while the modality-specific parts are optimized for self-reconstruction, ensuring effective representation learning.

\subsection{Remote Sensing-based} 
In recent years, earth observation via remote sensing has experienced a significant increase in data volume, posing challenges in managing and extracting pertinent information. To enhance search capabilities with greater expressiveness and flexibility, Psomas~\textit{et al.}~\cite{weicom} introduce the concept of CIR into remote sensing and build a benchmark dataset. As a pioneer study, it focuses on the attribute-based CIR, which aims to retrieve target images that share the same class(es) with the given reference image and possess the desired attribute(s) specified by the text description. This work designs a training-free method, WEICOM, which introduces a modality control parameter for balancing the importance of normalized image-oriented and text-oriented similarities. Both pretrained CLIP and RemoteCLIP~\cite{liu2024remoteclip} are explored for feature embedding. Beyond this work, Wang~\textit{et al.}~\cite{shf} study the remote sensing image retrieval with natural language-based text feedback. Recognizing that previous studies on CIR primarily focused on intrinsic attributes of target objects, while neglecting crucial extrinsic information such as spatial relationships in the remote sensing domain, Wang~\textit{et al.}~\cite{shf} propose a scene graph-aware hierarchical fusion network (SHF). This approach incorporates remote sensing image scene graphs as supplementary input data for enhancing structured image representation learning and target image retrieval. SHF employs a two-stage multimodal information fusion process. In the first stage, it fuses features from the remote sensing image scene graph and the remote sensing reference image at multiple levels to comprehensively capture the visual content. In the second stage, the modification text features are further fused with the final scene features from the first stage using a content modulator and a style modulator, which effectively capture and apply content and style changes.

\subsection{Dialog-based} 
The task of dialog-based CIR is proposed to address the challenge that users often struggle to initially express their intentions clearly or provide detailed descriptions of their objects of interest. Unlike traditional single-turn CIR, it allows users to iteratively refine their queries until they find a satisfactory item. This further necessitates models to integrate historical retrieval data with the current query to effectively locate the target image. 
Towards this end, DIIR~\cite{guo2018dialog} frames the task as a reinforcement learning problem, where the dialog system is rewarded for enhancing the rank of the target image within each dialog turn. Instead of using a real user to interact and train the dialog system, it introduces a user simulator based on the existing relative image captioning model \textit{Show, Attend, and Tell}~\cite{xu2015show}. The user simulator is trained on a newly collected dataset with Amazon Mechanical Turk, where both discriminative and relative captions are manually annotated within a shopping chatting scenario. Notably, since the data annotation for training a multi-turn simulator is quite expensive, the work only explored a single-turn user simulator. 
To enrich the decision-making process for retrieving the target image, CFIR~\cite{cfir2021} designs a three-way RNN-based framework, which evaluates the matching score of each candidate target image from three perspectives: 1) visual similarity between the composed query and the candidate image, 2) alignment between the reference-target image difference representation and the textual feedback representation; and 3) alignment between attribute representations of the candidate target image and the textual feedback representation. 
Meanwhile, CFIR contributes a large-scale multi-turn dataset, named Multi-turn FashionIQ, from the FashionIQ~\cite{wu2021fiq} by integrating multiple single-turn triplets. 
Inspired by the exceptional ability of Neural Turing Machines (NTMs)~\cite{graves2014neural} in handling complex long-term relationships, FashionNTM~\cite{fashionntm} introduces a Cascaded Memory Neural Turing Machine (CM-NTM) for multi-turn fashion image retrieval. Comprising a controller, individual read/write heads, and memory blocks, CM-NTM sequentially processes multi-turn queries with a cascaded mechanism, where each block's input comprises both the current turn's query feature and the output from the preceding block, supporting modeling the intricate relationships within input sequences. 
Beyond previous studies, IRR~\cite{irr} proposes a generative conversational composed retrieval framework, which formulates the task as a multimodal token sequence with alternating reference images and modification texts in the historical turns. IRR aims to autoregressively predict the target image feature based on the historical session data with the GPT decoder~\cite{radford2018improving}. 
Recently, LLM4MS~\cite{llm4ms} and Fashion-GPT~\cite{fashiongpt}  integrate LLMs into the retrieval system to realize dialog-based CIR. Specifically, LLM4MS employs a Q-former in BLIP2~\cite{li2023blip2} to translate image information into textual pseudo-tokens and then adopt an LLM T5~\cite{raffel2020exploring} to take into account all query information for retrieving the target image. To adapt the T5 model for this specific task while preserving its knowledge, LLM4MS employs LoRA~\cite{hu2021lora} techniques on the query and value matrices of all self-attention and cross-attention layers within T5 for efficient fine-tuning. Towards building a commercial fashion retrieval system, Fashion-GPT~\cite{fashiongpt} integrates ChatGPT with a pool of retrieval models in the fashion domain for handling users' diverse retrieval demands.

\subsection{Video-based} 
Composed video retrieval (CoVR) aims to retrieve target videos for a given reference image/video and a modification text. Ventura~\textit{et al.}~\cite{Covr,CoVR2} first introduce this task and develop models, named CoVR-BLIP and CoVR-BLIP2, by adapting BLIP and BLIP2 to the CoVR tasks, respectively. Since each video can be represented by sampled frames, these models are able to address both CIR and CoVR simultaneously. This work also contributes a dataset named WebVid-CoVR with $1.6$ million triplets by leveraging an LLM to automatically generate the modification text for two similar videos, identified by similar captions.
One key issue of this dataset is that its sample modifications are mainly regarding static color/shape/object changes, which do not need temporal understanding. To address this issue, Hummel~\textit{et al.}~\cite{tfrcvr} introduce EgoCVR, a manually curated dataset with $2,295$ queries. Additionally, they present a train-free method TF-CVR similar to the previous CIR model CIReVL~\cite{cirevl}. TF-CVR involves a video captioning model to generate a caption for the reference video. Then it uses an LLM to combine the video caption and modification text to obtain the target video caption and employs an existing text-to-video model to perform the cross-modal retrieval of target videos. To avoid selecting semantically similar but visually unrelated videos, they introduce a similarity-based filtering strategy to first narrow the scope of candidate videos for TF-CVR to rank.
To mine the rich query-specific context for promoting the target video retrieval, ECDE~\cite{ecde} introduces a novel CoVR framework. This framework uses detailed language descriptions of the given reference video derived from a multi-modal conversation model as the additional input to explicitly encode query-specific contextual information. For discriminative embedding learning, ECDE not only aligns the joint multimodal embedding of the input query with the conventional visual embedding of the target video, but also aligns that with the text embedding and vision-text joint embedding of the target video, respectively. 
\label{sec: exp}
\section{Benchmarks and Experiments}
In this section, we first present the public datasets related to CIR, and then provide experimental results and analyses of representative methods. 

\begin{table*}[h!]
\centering
\caption{\textbf{Statistics of datasets for composed image retrieval and its related tasks.}}
\label{tab:dataset_summary}
 \resizebox{12cm}{!}{
\begin{tabular}{l|c|c|c|c}
    \hline
    \textbf{Dataset} & \textbf{Data Type}& \textbf{Vision Scale} & \textbf{Triplet Scale} & \textbf{Triplet Construction} \\
    \hline \hline
    
    \multicolumn{5}{c}{\textit{\textcolor{gray}{Datasets for Composed Image Retrieval}}} \\
    
    FashionIQ~\cite{wu2021fiq} & image+text &  77.6K &  30.1K & Human Annotation \\ 
    Shoes~\cite{berg2010automatic} & image+text &  14.7K &  10.7K & Human Annotation \\
    CIRR~\cite{liu2021CIRPLANT} & image+text &  21.6K &  36.5K & Human Annotation \\
    B2W~\cite{forbes2019b2w} & image+text &  3.5K &  16.1K & Human Annotation \\
    CIRCO~\cite{searle} & image+text &  120K  &  1.0K & Human Annotation \\
    GeneCIS~\cite{genecis} & image+text & 33.3K &  8.0K & Human Annotation \\
    Fashion200K~\cite{Han2017fashion} & image+text &  200K &  205K & Template-base Generation \\
    MIT-States~\cite{Phi2015discover} & image+text &  53K & - & Template-base Generation \\
    CSS~\cite{vo2019tirg} & image+text & - &  32K & Template-base Generation \\
    SynthTriplets18M~\cite{compodiff} & image+text & - &  18.8M & Template-base Generation \\
    LaSCo~\cite{levy2024case} & image+text &  121.5K &  389.3K & LLM-base Generation \\

    \cdashline{1-5}
    \multicolumn{5}{c}{\textit{\textcolor{gray}{Datasets for Related Tasks of Composed Image Retrieval}}} \\

    Shopping100k~\cite{EMASL} & image+attributes &  101K &  1.1M & Template-base Generation \\
    WebVid-CoVR~\cite{ventura2024covr} & video+text &  130.8K &  1.6M & LLM-base Generation \\
    FS-COCO~\cite{chowdhury2022fscoco} & sketch+image+text &  10K &  10K & Template-base Generation \\
    SketchyCOCO~\cite{gao2020sketchcoco} & sketch+image+text &  14K &  14K & Template-base Generation \\
    CSTBIR~\cite{stnet} & sketch+image+text &  108K &  2M & Template-base Generation \\
    PATTERNCOM~\cite{psomas2024cir4rs} & image+text&  30K &  21K & Template-base Generation\\
    Airplane, Tennis, and WHIRT~\cite{shf} & image+scene graph+text&  7.7K &  8.7K & Human Annotation \\
    Multi-turn FashionIQ~\cite{cfir2021} & image+text & 13.6K & 11.5K & Human Annotation \\
    
    \hline
    \multicolumn{5}{l}{\small *Vision scale means the scale of images/sketch-image pairs/videos. The triplet scale in Multi-turn FashionIQ refers to the number of sessions.}
\end{tabular}
 }
\end{table*}

\subsection{Datasets.}
The statistics of datasets for the CIR and its related tasks are summarized in Table~\ref{tab:dataset_summary}.

\textbf{FashionIQ.} 
The FashionIQ dataset~\cite{wu2021fiq} is a natural language-based interactive fashion retrieval dataset, crawled from \textit{Amazon.com}. It provides human-generated captions that distinguish similar pairs of garment images together. The fashion items within the dataset belong to three categories: dress, shirt, and top\&tee. It contains $\sim77.6$K images and $\sim30.1$K triplets, with $\sim46.6$K images and $\sim18$K triplets in the training set, $\sim15.5$K images and $\sim6$K triplets in the validation set, and $\sim15.5$K images and $\sim6$K triplets in the test set. While being a challenge dataset, the test set is not publicly accessible.
Notably, FashionIQ comprises two evaluation protocols: the VAL-Split~\cite{chen2020val} and the Original-Split~\cite{wu2021fiq}. The VAL-Split is introduced by the early-stage CIR study, which constructs the candidate image set for testing based on the union of the reference images and target images in all the triplets of the validation set. The Original-Split has recently been adopted, and it directly uses the original candidate image set provided by the FashionIQ dataset for testing. 


\textbf{Shoes.}
The Shoes dataset~\cite{berg2010automatic} is originally collected from \textit{like.com} for the attribute discovery task and further developed by~\cite{guo2018dialog} with relative caption annotations for dialog-based interactive retrieval. The annotations are gathered via human annotation using an interactive interface, which allows for fine-grained attribute descriptions. The dataset includes categories such as boots, sneakers, high heels, clogs, pumps, rain boots, flats, stilettos, wedding shoes, and athletic shoes. Overall, the dataset comprises $\sim14.7$K images and $\sim10.7$K triplets, with $10$K images and $\sim9$K triplets for training, and $\sim4.7$K images and $\sim1.7$K triplets for testing. 

\textbf{CIRR.} The CIRR dataset, introduced by Liu~\textit{et al.}~\cite{liu2021CIRPLANT}, is an open-domain dataset constructed using  $\sim21.5$K images sourced from the natural language reasoning dataset $\text{NLVR}^2$~\cite{suhr2019nvlr}. CIRR comprises $\sim36.5$K triplets divided into training, validation, and test sets with an allocation ratio of $8:1:1$. To alleviate the false negative issue, CIRR first clusters similar images into subsets based on their visual similarity before the reference-target image pairs construction. Then, during the subsequent process of annotating modification text for reference-target image pairs, the semantics of the modification text must differentiate the target image from other similar images within the same subset. Specifically, given that each subset contains samples with a high visual similarity to the target image, testing retrieval on this subset places greater demands on the model's discriminative ability.

\textbf{B2W.}
The Birds-to-Words (B2W) dataset~\cite{forbes2019b2w} consists of images of birds sourced from \textit{iNaturalist}, accompanied by paragraphs written by humans to describe the differences between pairs of images. The dataset contains approximately $\sim3.5$K images and $\sim16.1$K triplets. Notably, each text description is relatively detailed, with an average length of $31.38$ words, providing rich insights into the subtle variations across bird images.

\textbf{CIRCO.} The CIRCO dataset~\cite{searle} is an open-domain dataset developed from the COCO 2017 unlabeled set~\cite{lin2014coco} to address the false negative issues prevalent in existing datasets. Unlike typical CIR datasets, CIRCO includes multiple target images per query, thus significantly reducing the occurrence of false negatives and establishing it as the first CIR dataset with multiple ground-truth target images. Given that CIRCO is specifically designed for evaluating zero-shot cross-image retrieval models, it comprises $1,020$ queries that are partitioned into a validation set and a test set. Specifically, $220$ queries are allocated for validation purposes, while the remaining $800$ for testing. Each query includes a reference image, modification text, and an average of $4.53$ ground truth target images. Utilizing all $120$K images of COCO as the index set, CIRCO provides a vastly larger number of distractors compared to the $2$K images in the CIRR test set.

\textbf{GeneCIS.} 
GeneCIS~\cite{genecis} is an open-domain CIR dataset that serves as a benchmark for evaluating conditional similarity tasks. This dataset includes four subsets: Focus Attribute, Change Attribute, Focus Object, and Change Object, representing four different tasks. Among these, Focus Object and Change Object are constructed based on the COCO~\cite{lin2014coco} dataset, while Focus Attribute and Change Attribute are constructed based on the VAW~\cite{pham2021learning} dataset. Unlike other open-domain datasets, CIRR and CIRCO, which provide modification text, GeneCIS provides a single object name or attribute as the retrieval condition. To reduce the impact of false negatives, a gallery was selected for each triplet as the retrieval candidate set, with gallery sizes ranging from 10 to 15 images. This dataset consists of $\sim8$K triplets.

\textbf{Fashion200K.} The Fashion200K dataset, collected by Han~\textit{et al.}~\cite{Han2017fashion}, comprises  $\sim200$K clothing images, which are categorized into five types: dress, top, pants, skirt, and jacket. Each image comes with a compact attribute-like product description, such as ``black biker jacket''. The dataset is divided into three parts: $\sim172$K images for the training set, $\sim12$K images for the validation set, and $\sim25$K images for the test set. To construct triplets suitable for the CIR task, existing works~\cite{vo2019tirg, wen2021clvcnet} first create image pairs by identifying only one-word differences in their descriptions. The modification text is generated using templates that incorporate the differing words, such as ``replace red with green.'' Based on this construction method, there are $\sim172$K triplets available for training and $\sim33$K triplets for evaluation. 

\textbf{MIT-States.} The MIT-States dataset~\cite{Phi2015discover} features a diverse collection of objects, scenes, and materials in various transformed states. It contains $\sim53$K images, in which each image is tagged with an adjective or state label and a noun or object label (\textit{e.g.}, ``new camera'' or ``cooked beef''). The dataset encompasses $245$ nouns and $115$ adjectives, with each noun being modified by roughly $9$ different adjectives on average. Following~\cite{vo2019tirg}, image pairs that share the same object labels but with different attributes or states can be selected as the reference and target images, and the different attributes or states serve as the modification text. To evaluate the model's capacity for handling unseen objects, $49$ nouns are used for the test and the rest is for training.

\textbf{CSS.} The CSS dataset~\cite{vo2019tirg} is constructed using the CLEVR toolkit~\cite{johnson2017clevr} to generate synthesized images in a $3$-by-$3$ grid scene, showcasing objects with variations in Color, Shape, and Size. Each image is available in both a simplified $2$D blob version and a detailed $3$D rendered version. The dataset comprises $\sim16$K queries for training and $\sim16$K queries for test. Each query is of a reference image ($2$D or $3$D) and a modification, and the target image. Notably, modification texts fall into three categories: adding, removing, or changing object attributes. Examples include ``add red cube'' or ``remove yellow sphere''.

\textbf{SynthTriplets18M.} SynthTriplets18M~\cite{compodiff} is a large-scale dataset specifically designed for the CIR task. Distinct from datasets relying on human annotation, SynthTriplets18M uses diffusion models to automatically create $\sim18$M triplets consisting of reference images, modification instructions, and target images. Adopting the approach of Instruct Pix2Pix~\cite{brooks2023pix}, the dataset initially creates caption triplets by modifying reference captions with specific instructions and then converts these triplets into image-based triplets using diffusion models, resulting in highly diverse and realistic data. The dataset encompasses a broad spectrum of keywords, enhancing its suitability for open-domain CIR tasks, and provides rich textual prompts such as ``replace ${source} with ${target}.'' This automated generation process enables the creation of rare and diverse image triplets that might not commonly appear in reality.

\textbf{LaSCo.} LaSCo~\cite{levy2024case} is a large-scale, open-domain dataset consisting of natural images, specifically designed for the CIR task. Created with minimal human effort by leveraging the VQA $2.0$ dataset~\cite{goyal2017vqa}, LaSCo leverages ``complementary'' image pairs, which are similar images that yield different answers to the same question, and transforms these question-answer pairs into valid transition texts using GPT-$3$. By exploiting transition symmetry, the dataset has amassed $\sim121.5$K images and $\sim389.3$K image pairs, which are then organized into triplets.

\textbf{Shopping100k.} The Shopping100k dataset~\cite{EMASL} consists of $\sim101.0$K pure clothing images that are characterized by $12$ attributes with $151$ possible attribute values. Following~\cite{adde}, pairs of images that differ by only one or two attributes are utilized as reference and target images. These differing attributes serve as modification attributes, facilitating the construction of triplet data for attribute-based CIR tasks. This approach results in $\sim954.1$K training triplets and $\sim118.3$K testing triplets, forming a robust foundation for developing and evaluating CIR models focused on nuanced fashion attributes.

\textbf{WebVid-CoVR.}
WebVid-CoVR~\cite{ventura2024covr} is a large-scale dataset designed for CoVR, containing $\sim1.6$M triplets. Each video in the dataset averages $16.8$ seconds in duration, and the modification texts consist of roughly $4.8$ words. With each target video linked to about $12.7$ triplets, the dataset offers rich contextual variations essential for effectively training CoVR models. To ensure robust evaluation, they further introduce a meticulously curated test set known as WebVid-CoVR-Test~\cite{ventura2024covr}, which is manually annotated and consists of $\sim2.5$K triplets.



\textbf{FS-COCO.} FS-COCO~\cite{chowdhury2022fscoco} is a large-scale dataset of freehand scene sketches paired with textual descriptions and corresponding images, designed to advance research in fine-grained scene sketch understanding. It consists of $\sim10$K unique sketches drawn by non-experts, with $\sim7$K for training and $\sim3$K for testing, each matched with a photo from the MS-COCO dataset~\cite{lin2014microsoft} and a descriptive caption. 
FS-COCO includes over $90$ object categories from the COCO-stuff~\cite{caesar2018coco} and provides temporal order information of strokes, which enables detailed studies on scene abstraction and the salience of early vs. late strokes. This dataset serves as a benchmark for fine-grained image retrieval, sketch-based captioning, and understanding the complementary information between sketches and text.

\textbf{SketchyCOCO.} SketchyCOCO~\cite{gao2020sketchcoco} is a large-scale composite dataset tailored for the task of automatic image generation from scene-level freehand sketches. Built on the COCO-stuff dataset~\cite{caesar2018coco}, it includes $\sim14$K unique scene-level sketches paired with corresponding images and textual descriptions, organized into $\sim14$K  sketch-text-image triplets, with $80$\% for training and the remaining $20$\% for testing. Additionally, the dataset includes $\sim20$K triplet examples of foreground sketches, images, and edge maps covering $14$ classes, along with $\sim27$K background sketch-image pairs covering $3$ classes. This layered structure facilitates detailed studies in scene-level sketch-based generation and provides five-tuple data for comprehensive training in both foreground and background synthesis tasks.

\textbf{CSTBIR.} The Composite Sketch+Text Based Image Retrieval (CSTBIR) dataset~\cite{stnet} is a multimodal dataset specifically designed for image retrieval using sketches and partial text descriptions. It includes $\sim108$K natural scene images and $\sim2$M annotated triplets, each containing a reference sketch, a partial text description, and a target image. The natural images and text descriptions are sourced from the Visual Genome dataset~\cite{krishna2017visual}, while the sketches are hand-drawn from the Quick, Draw! dataset~\cite{HaE18}. By intersecting object categories between Visual Genome and Quick, Draw!, the dataset contains $258$ intersecting object classes and is divided into training, validation, and testing sets aligned with Visual Genome’s splits. The training set comprises $\sim97$K images, $\sim484$K sketches, and $\sim1.89$M queries. Additionally, CSTBIR includes three distinct test sets: Test-$1$K, Test-$5$K, and an Open-Category set. Among them, only the Open-Category test set is designed to evaluate model performance on novel object categories not present during training, which contains $70$ novel object categories (of which $50$ are “difficult-to-name”) and corresponding sketches.

\textbf{PATTERNCOM.} PATTERNCOM~\cite{psomas2024cir4rs} is a new benchmark designed for evaluating remote sensing CIR methods, based on the PatternNet dataset~\cite{zhou2018patternnet}, which is a large-scale, high-resolution remote sensing image collection. PATTERNCOM focuses on selected classes from PatternNet, incorporating query images and corresponding text descriptions that define relevant attributes for each class. For example, the ``swimming pools'' category includes text queries specifying shapes such as ``rectangular,'' ``oval,'' and ``kidney-shaped.'' The dataset encompasses six attributes, with each attribute linked to up to four different classes and two to five values per class. Overall, PATTERNCOM contains over $21$K queries, with positive matches ranging from $2$ to $1,345$ per query.

\textbf{Airplane, Tennis, and WHIRT.} The Airplane, Tennis, and WHIRT datasets~\cite{shf} are curated for remote sensing CIR. 
Each dataset is organized in terms of quintets, consisting of a reference RS image and its scene graph, a target RS image and its scene graph, and a pair of modifier sentences. The Airplane dataset comprises $1,600$ remote sensing images from UCM~\cite{yang2010bag}, PatternNet~\cite{zhou2018patternnet}, and NWPU-RESISC45~\cite{cheng2017remote}, along with $3,461$ pairs of modifier sentences that describe differences in airplane attributes and spatial relationships between airplanes and other objects. The Tennis dataset includes $1,200$ images featuring tennis courts and $1,924$ manually annotated modifier sentence pairs. It emphasizes target and non-target spatial relationships while ignoring relationships between non-target objects. The WHIRT dataset is the most extensive, consisting of $4,940$ images from WHDLD~\cite{shao2018performance} and $3,344$ reference-target image pairs. Its scene graphs provide comprehensive details on object attributes and spatial relationships, accommodating complex remote sensing scenarios. Each dataset features high-quality annotations by domain experts, serving as robust benchmarks for advanced retrieval tasks in intricate remote sensing environments. 

\textbf{Multi-turn FashionIQ.} Multi-turn FashionIQ~\cite{cfir2021} is an extension of the original FashionIQ dataset~\cite{wu2021fiq}, designed to model user interactions in a multi-turn setting for fashion product retrieval. It contains $\sim11.5$K sessions structured as multi-turn interaction, across three clothing types: dress, shirt, and top\&tee. Each session consists of multiple reference images, modification texts, and a target image, with turns ranging from $3$ to $5$. The sessions are constructed by linking single-turn triplets from FashionIQ, matching the target image of one triplet to the reference image of another, thus forming coherent multi-turn sequences. Additionally, the dataset expands the original attribute data to ensure comprehensive coverage, associating each target image with attributes like texture, fabric, shape, part, and style. 

\subsection{Metric.}
\textbf{Recall.} Recall is a widely used metric in CIR to evaluate the effectiveness of retrieval systems. It is often denoted as Recall@$k$ (R@$k$), measuring the proportion of queries for which the correct target image is retrieved within the top $k$ results. Recall@$k$ can be defined by the following formula:
\begin{equation}
    {\text{Recall}@k} = \frac{1}{Q} \sum_{q=1}^{Q} \frac{| \mathcal{R}_q \cap \mathcal{D}_q^k |}{| \mathcal{R}_q |} ,
\end{equation}
where $Q$ denotes the total number of queries, $\mathcal{R}_q$ is the set of all relevant target images for each query $q$, $\mathcal{D}_q^k$ is the set of the top $k$ retrieved items for query $q$, $| \mathcal{R}_q \cap \mathcal{D}_q^k|$ represents the number of target images found in the top $k$ results, $| \mathcal{R}_q |$ represents the total number of target images for query $q$. Notably, in most existing CIR datasets, each query typically corresponds to only one target image, \textit{i.e.}, $| \mathcal{R}_q |=1$, except CIRCO. Additionally, CIRR~\cite{liu2021CIRPLANT} further defines Recall\(_{subset}@k\) to assess how frequently the desired target image appears in the top $k$ results when considering only a specific subset of images. 

\textbf{Mean Average Precision.}
Mean Average Precision at $k$ (mAP@$k$) is a crucial metric for evaluating retrieval systems, particularly in cases where there are multiple relevant items. Initially employed in CIRCO~\cite{searle}, this metric integrates precision across various ranks to yield a single, averaged indicator of the system's effectiveness in retrieving relevant items. The formula for mAP@$k$ is given by:
\begin{equation}
    \text{mAP}@k = \frac{1}{Q} \sum_{q=1}^{Q} \frac{1}{\min(k, \mathcal{R}_q)} \sum_{k=1}^{k} P@k \cdot \text{rel}@k,
\end{equation}
where P@$k$ is the precision at rank $k$, rel@$k$ is a relevance function. The relevance function is an indicator function that equals $1$ if the image at rank $k$ is labeled as positive and equals $0$ otherwise.

\begin{table*}
    \scriptsize
    \centering
    \caption{
   Performance comparison among supervised composed image retrieval models on FashionIQ (VAL split).}
\begin{tabular}{l|cc|cc|cc|cc|c}
    \hline 
    \multirow{2}{*}{Model} & \multicolumn{2}{c|}{Dresses} & \multicolumn{2}{c|}{Shirts} & \multicolumn{2}{c|} {Tops\&Tees} & \multicolumn{2}{c|}{Average} & \multirow{2}{*}{Avg.} \\ \cline{2-9}
    &  R@$10$ & R@$50$ & R@$10$ & R@$50$ & R@$10$ & R@$50$ & R@$10$ & R@$50$ & \\
    \hline \hline
    
    \multicolumn{10}{c}{\textit{\textcolor{gray}{Traditional Encoder-based Methods}}} \\
    
    LSC4TCIR~\cite{chawla2021lsc4cir} \footnotesize{\textcolor{gray}{(CVPRW'21)}} & $19.33 $ & $43.52$ & $14.47$ & $35.47$ & $19.73$ & $44.56$ & $17.84$ & $41.18$ &$29.51$ \\
    CIRPLANT~\cite{liu2021CIRPLANT} \footnotesize{\textcolor{gray}{(ICCV'21)}} & $17.45$ & $40.41$ & $17.53$ & $38.81$ & $21.64$ & $45.38$ & $18.87$ & $41.53$ & $30.20$ \\
    VAL~\cite{chen2020val} \footnotesize{\textcolor{gray}{(CVPR'20)}} & $21.12$ & $42.19$ & $21.03$ & $43.44$ & $25.64$ & $49.49$ & $22.60$ & $45.04$ & $33.82$ \\
    DATIR~\cite{gu2021datir} \footnotesize{\textcolor{gray}{(MM'21)}} & $21.90$ & $43.80$ & $21.90$ & $43.70$ & $27.20$ & $51.60$ & $23.70$ & $46.40$ & $35.00$ \\
    SynthTripletGAN~\cite{tautkute2021Synth} \footnotesize{\textcolor{gray}{(Arxiv'21)}} & $22.60$ & $45.10$ & $20.50$ & $44.08$ & $28.01$ & $52.10$ & $23.70$ & $47.09$ & $35.40$ \\
\hline
    \end{tabular}
    \label{tab:supervised_CIR_exp_fashioniq_val}
\end{table*}

\begin{table*}
    \scriptsize
    \centering
    \caption{
   Performance comparison among supervised composed image retrieval models on FashionIQ (original split).}
    \begin{tabular}{l|cc|cc|cc|cc|c}
    \hline 
    \multirow{2}{*}{Model} & \multicolumn{2}{c|}{Dresses} & \multicolumn{2}{c|}{Shirts} & \multicolumn{2}{c|} {Tops\&Tees} & \multicolumn{2}{c|}{Average} & \multirow{2}{*}{Avg.} \\ \cline{2-9}
    &  R@$10$ & R@$50$ & R@$10$ & R@$50$ & R@$10$ & R@$50$ & R@$10$ & R@$50$ & \\
    \hline \hline
    
    \multicolumn{10}{c}{\textit{\textcolor{gray}{Traditional Encoder-based Methods}}} \\
    
    JVSM~\cite{chen2020jvsm} \footnotesize{\textcolor{gray}{(ECCV'20)}} & $10.70 $ & $25.90$ & $12.00$ & $27.10$ & $13.00$ & $26.90$ & $11.90$ & $26.63$ &$19.27$ \\
    TIRG~\cite{vo2019tirg} \footnotesize{\textcolor{gray}{(CVPR'19)}} & $14.13$ & $34.61$ & $13.10$ & $30.91$ & $14.79$ & $34.37$ & $14.01$ & $33.30$ &$23.66$ \\
    AlRet-small~\cite{xu2024alret} \footnotesize{\textcolor{gray}{(TMM'24)}} & $27.34$ & $53.42$ & $21.30$ & $43.08$ & $29.07$ & $54.21$ & $25.90$ & $50.24$ &$38.07$ \\
    ARTEMIS~\cite{delmas2022artemis} \footnotesize{\textcolor{gray}{(ICLR'22)}} & $27.16$ & $52.40$ & $21.78$ & $43.64$ & $29.20$ & $54.83$ & $26.05$ & $50.29$ & $38.17$ \\
    FashionVLP~\cite{goenka2022fashionvlp} \small{\textcolor{gray}{(CVPR'22)}} & $26.77$ & $53.20$ & $22.67$ & $46.22$ & $28.51$ & $57.47$ & $25.98$ & $52.30$ & $39.14 $ \\
\hline
    \end{tabular}
    \label{tab:supervised_CIR_exp_fashioniq_ori}
\end{table*}

\begin{table*}
    \scriptsize
    \centering
    \caption{Performance comparison among supervised composed image retrieval models on Fashion200K and Shoes.}
    \begin{tabular}{l|ccc|c|ccc|c}
    \hline 
    \multirow{2}{*}{Model} & \multicolumn{4}{c|}{Fashion200K} & \multicolumn{4}{c}{Shoes}  \\
    \cline{2-9}
    & R@$1$ & R@$10$ & R@$50$ & Avg. & R@$1$ & R@$10$ & R@$50$ & Avg.\\
    \hline \hline 
    
    \multicolumn{9}{c}{\textit{\textcolor{gray}{Traditional Encoder-based Methods}}} \\
    TIRG~\cite{vo2019tirg} \footnotesize{\textcolor{gray}{(CVPR'19)}} & $14.10$ & $42.50$ & $63.80$ &$40.13$ &$12.60$ &$45.45$ & $69.39$&$42.28$\\
    MGF~\cite{liu2021mgf} \footnotesize{\textcolor{gray}{(MMM'21)}} & $16.00$ & $44.60$ & $-$ &$-$ &$-$ &$-$ & $-$&$-$ \\
    TIRG+JPM(Tri)~\cite{yang2021jpm} \footnotesize{\textcolor{gray}{(MM'21)}} & $17.70$ & $44.70$ & $64.50$ &$42.30$ &$-$ &$-$ & $-$&$-$\\
    JAMMA~\cite{zhang2020jamma} \footnotesize{\textcolor{gray}{(MM'20)}} & $17.34$ & $45.28$ & $65.65$ &$42.76$ &$-$ &$-$ & $-$&$-$\\
    DCNet~\cite{kim2021dcnet} \footnotesize{\textcolor{gray}{(AAAI'21)}} & $-$ & $46.89$ & $67.56$ &$-$ & $-$ & $53.82$ & $79.33$ & $-$ \\
\hline
    \end{tabular}
    \label{tab:supervised_CIR_exp_fashion200k_shoes_css}
\end{table*}

\begin{table*}
    \scriptsize
    \centering
    \caption{Performance comparison among supervised composed image retrieval models on CIRR. Avg means the average of R@$5$ and R$_{subset}$@$1$.}
    \begin{tabular}{l|cccc|ccc|c}
    \hline 
    \multirow{2}{*}{Method} &\multicolumn{4}{c|}{\textbf{R@$k$}} &\multicolumn{3}{c|}{\textbf{R$_{subset}$@$k$}} & \multirow{2}{*}{Avg} \\ \cline{2-8}
    & $k=1$ & $k=5$ & $k=10$ & $k=50$ & $k=1$ & $k=2$ & $k=3$ \\
    \hline \hline 

    \multicolumn{9}{c}{\textit{\textcolor{gray}{Traditional Encoder-based Methods}}} \\
    
    TIRG~\cite{vo2019tirg} \footnotesize{\textcolor{gray}{(CVPR'19)}} & $14.61$ & $48.37$ & $64.08$ & $90.03$ & $22.67$ & $44.97$ & $65.14$ & $35.52$\\
    ARTEMIS~\cite{delmas2022artemis} \footnotesize{\textcolor{gray}{(ICLR'22)}} & $16.96$ & $46.10$ & $61.31$ & $87.73$ & $39.99$ & $62.20$ & $75.67$ & $43.05$ \\
    CIRPLANT~\cite{liu2021CIRPLANT} \footnotesize{\textcolor{gray}{(ICCV'21)}} & $19.55$ & $52.55$ & $68.39$ & $92.38$ & $39.20$ & $63.03$ & $79.49$ & $45.88$ \\
    MCEM~\cite{zhang2024mcem} \footnotesize{\textcolor{gray}{(TIP'24)}} & $17.48$ & $46.13$ & $62.17$ & $88.91$ & $-$ & $-$ & $-$ & $-$ \\
    NEUCORE~\cite{zhao2024neucore} \footnotesize{\textcolor{gray}{(NIPSW'23)}} & $18.46$ & $49.40$ & $63.57$ & $89.35$ & $44.27$ & $67.06$ & $78.92$ & $46.84$ \\
\hline
    \end{tabular}
    \label{tab:supervised_CIR_exp_cirr}
\end{table*}

\begin{table*}
    \scriptsize
    \centering
    \caption{Performance comparison among supervised composed image retrieval models on MIT-States and CSS.}
    \begin{tabular}{l|ccc|c|cc}
    \hline 
    \multirow{2}{*}{Model} & \multicolumn{4}{c|}{MIT-States} & \multicolumn{2}{c}{CSS}  \\
    \cline{2-7}
    & R@$1$ & R@$5$ & R@$10$ & Avg. & R@$1$(3D-to-3D) & R@$1$(2D-to-3D) \\
    \hline \hline 

    \multicolumn{7}{c}{\textit{\textcolor{gray}{Traditional Encoder-based Methods}}} \\
    TIRG~\cite{vo2019tirg} \footnotesize{\textcolor{gray}{(CVPR'19)}} & $12.20$ & $31.90$ & $43.10$ &$29.07$ & $73.70$ & $46.60$\\
    MGF~\cite{liu2021mgf} \footnotesize{\textcolor{gray}{(MMM'21)}} & $13.10$ & $-$ & $43.60$ & $-$ & $74.20$ & $62.80$\\
    TIS~\cite{zhang2022tis} \footnotesize{\textcolor{gray}{(TOMM'22)}} & $13.13$ & $31.94$ & $43.32$ & $29.46$ & $76.64$ & $48.02$\\
    TIRG+JPM(Tri)~\cite{yang2021jpm} \footnotesize{\textcolor{gray}{(MM'21)}} & $-$ & $-$& $-$ & $-$& $83.20$ & $-$\\
    TIRG+JPM(MSE)~\cite{yang2021jpm} \footnotesize{\textcolor{gray}{(MM'21)}} & $-$ & $-$& $-$ & $-$& $83.80$ & $-$\\
\hline
    \end{tabular}
    \label{tab:supervised_CIR_exp_mit_css}
\end{table*}

\begin{table*}
    \scriptsize
    \centering
    \caption{Performance comparison among zero-shot composed image retrieval models on FashionIQ (original split).}
    \begin{tabular}{l|c|cc|cc|cc|cc|c}
    \hline 
    \multirow{2}{*}{Model} & \multirow{2}{*}{Encoder} & \multicolumn{2}{c|}{Dresses} & \multicolumn{2}{c|}{Shirts} & \multicolumn{2}{c|} {Tops\&Tees} & \multicolumn{2}{c|}{Average} & \multirow{2}{*}{Avg.} \\ \cline{3-10}
    & & R@$10$ & R@$50$ & R@$10$ & R@$50$ & R@$10$ & R@$50$ & R@$10$ & R@$50$ & \\
    \hline \hline
    
    \multicolumn{11}{c}{\textit{\textcolor{gray}{Textual-inversion-based Methods}}}\\
    SEARLE~\cite{searle} \footnotesize{\textcolor{gray}{(ICCV'23)}} & CLIP-B & $18.54$ & $39.51$ & $24.44$ & $41.61$ & $25.70$ & $46.46$ & $22.89$ & $42.53$ & $32.71$ \\
    iSEARLE~\cite{isearle} \footnotesize{\textcolor{gray}{(Arxiv'24)}} & CLIP-B & $20.92$ & $42.19$ & $25.81$ & $43.52$ & $26.47$ & $48.70$ & $24.40$ & $44.80$ & $34.60$\\
    
    Pic2Word~\cite{pic2word} \footnotesize{\textcolor{gray}{(CVPR'23)}} & CLIP-L & $20.00 $ & $40.20$ & $26.20$ & $43.60$ & $27.90$ & $47.40$ & $24.70$ & $43.70$ &$34.22$\\
    SEARLE-XL~\cite{searle} \footnotesize{\textcolor{gray}{(ICCV'23)}} & CLIP-L & $20.48$ & $43.13$ & $26.89$ & $45.58$ & $29.32$ & $49.97$ & $25.56$ & $46.23$ & $35.90$ \\
    LinCIR~\cite{lincir} \footnotesize{\textcolor{gray}{(CVPR'24)}} & CLIP-L & $20.92$ & $42.44$ & $29.10$ & $46.81$ & $28.81$ & $50.18$ & $26.28$ & $46.49$ & $36.38$ \\
\hline
    \end{tabular}
    \label{tab:zs_CIR_exp_fashioniq_ori}
\end{table*}

\begin{table*}
    \scriptsize
    \centering
    \caption{Performance comparison among zero-shot composed image retrieval models on CIRR and CIRCO.}
    \begin{tabular}{l|c|cccc|ccc|cccc}
    \hline 
    \multirow{3}{*}{Model} & \multirow{3}{*}{Encoder} & \multicolumn{7}{c|}{CIRR} & \multicolumn{4}{c}{CIRCO}  \\ 
    \cline{3-13}
    && \multicolumn{4}{c|}{\textbf{R@$k$}} & \multicolumn{3}{c|}{\textbf{R$_{subset}$@$k$}} & \multicolumn{4}{c}{\textbf{mAP@$k$}}\\
    \cline{3-13}
    & & $k=1$ & $k=5$ & $k=10$ & $k=50$ & $k=1$ & $k=2$ & $k=3$ & $k=5$ & $k=10$ & $k=25$ & $k=50$ \\
    \hline \hline
    
    \multicolumn{13}{c}{\textit{\textcolor{gray}{Textual-inversion-based Methods}}}\\
    SEARLE~\cite{searle} \footnotesize{\textcolor{gray}{(ICCV'23)}} & CLIP-B & $24.00 $ & $53.42$ & $66.82$ & $89.78$ & $54.89$ & $76.60$ & $88.19$ & $9.35$ & $9.94$ & $11.13$ &$11.84$\\
    iSEARLE~\cite{isearle} \footnotesize{\textcolor{gray}{(Arxiv'24)}} & CLIP-B & $25.23$ & $55.69$ & $68.05$ & $90.82$ & $-$ & $-$ & $-$ & $10.58$ & $11.24$ & $12.51$ & $13.26$\\
    Pic2Word~\cite{pic2word} \footnotesize{\textcolor{gray}{(CVPR'23)}} & CLIP-L & $23.90 $ & $51.70$ & $65.30$ & $87.80$ & $-$ & $-$ & $-$ & $-$ & $-$ & $-$ &$-$\\
    SEARLE-XL~\cite{searle} \footnotesize{\textcolor{gray}{(ICCV'23)}} & CLIP-L & $24.24 $ & $52.48$ & $66.29$ & $88.84$ & $53.76$ & $75.01$ & $88.19$ & $11.68$ & $12.73$ & $14.33$ &$15.12$\\
\hline
    \end{tabular}
    \label{tab:zs_CIR_exp_cirr_circo_ori}
\end{table*}

\subsection{Experimental Results.}
In this subsection, we compare and analyze supervised CIR and ZS-CIR methods as reviewed above.

\subsubsection{Supervised Composed Image Retrieval.} 
To have an in-depth insight into the results of supervised CIR methods, we provide a comparison of their performance on various widely used datasets in Tables~\ref{tab:supervised_CIR_exp_fashioniq_ori}-~\ref{tab:supervised_CIR_exp_mit_css}. These datasets include FashionIQ, Fashion200k, MIT-States, CSS, Shoes, and CIRR. 
Notably, the FashionIQ dataset includes two evaluation protocols. However, some current approaches have erroneously intermixed results from different protocols during model comparison. To address this issue, we conduct a detailed inspection of the methods and their available source code. We then organize the comparisons for the VAL split and the original split separately to ensure fairness.
Furthermore, recognizing the significant impact of different encoders on model performance, we categorize the methods into two groups: those utilizing traditional encoders, such as ResNet and LSTM, and those employing VLP encoders (\textit{e.g.}, CLIP and BLIP) as the feature extraction backbones. 
From these tables, we obtain the following observations. 
1) VLP encoder-based methods generally achieve much better performance in comparison to traditional encoder-based methods. The reason is that VLP encoders are usually larger than traditional encoder methods. Moreover, VLP encoders are typically pre-trained on extensive corpora of image-text pairs through contrastive learning, thereby possessing excellent capabilities for multimodal alignment and cross-modal retrieval, which are crucial in the context of CIR. This can also be emphasized by methods that adopt the same fusion strategy but different types of encoders. As can be seen from Tables~\ref{tab:supervised_CIR_exp_fashioniq_val}-\ref{tab:supervised_CIR_exp_fashion200k_shoes_css}, AIRet-big with the VLP encoder attains significantly better performance than AIRet-small with a traditional encoder. Additionally, even when using the same type of traditional encoder or VLP encoder, the version with larger parameters typically delivers better performance. For example, LBF-big outperforms LBF-small (See Table~\ref{tab:supervised_CIR_exp_fashion200k_shoes_css}), and FashionERN-big surpasses FashionERN-small (See Table~\ref{tab:supervised_CIR_exp_fashioniq_ori}).
These results highlight that the choice of image and text encoders plays a pivotal role in the context of CIR, often surpassing the importance of image-text fusion and target matching method design.
2) Regarding image-text fusion, various strategies have demonstrated strong performance. Here, we summarize the top-performing methods for the FashionIQ-VAL, FashionIQ-ori, Fashion200k, Shoes, CIRR, MIT-States, and CSS datasets, respectively. Notably, to ensure a fair comparison, only prototype methods are evaluated, with additional modules such as data augmentation and reranking disabled.
The results show that DQU-CIR (MLP-based), SDQUR (Cross-attention-based), SPRC (Self-attention-based), and GSCMR (Graph-attention-based) achieve the best performance across these datasets. This indicates that while the image-text fusion strategy is a critical component of CIR methods, it is not the sole determinant of final performance. Moreover, it remains challenging to identify a universally optimal fusion strategy.
3) Models that integrate additional target matching techniques or data augmentation modules consistently demonstrate superior performance in comparison to their original counterparts. For example, SPRC-VQA, which employs visual question answering to re-rank the retrieval list, outperforms SPRC, as demonstrated in Table \ref{tab:supervised_CIR_exp_fashioniq_ori} and Table \ref{tab:supervised_CIR_exp_cirr}. LIMN+, by leveraging augmented pseudo-triplets for iterative training, attains better outcomes than LIMN shown in Table \ref{tab:supervised_CIR_exp_fashioniq_val} and Table \ref{tab:supervised_CIR_exp_fashion200k_shoes_css}. Additionally, ComposeAE+GA, which adopts the gradient augmentation regularization approach to achieve the dataset augmentation effect and mitigate overfitting, surpasses ComposeAE, as can be seen from Table \ref{tab:supervised_CIR_exp_fashion200k_shoes_css} and Table \ref{tab:supervised_CIR_exp_mit_css}. These examples emphasize the critical significance of further exploiting advanced target matching strategies and implementing dataset augmentation techniques to bolster the model's generalization capabilities.


\subsubsection{Zero-shot Composed Image Retrieval.} ZS-CIR models, including textual-inversion-based, pseudo-triplet-based, and training-free models, are evaluated on the FashionIQ, CIRR, and CIRCO datasets. We directly summarize their experimental results from the corresponding papers in Table~\ref{tab:zs_CIR_exp_fashioniq_ori} and Table~\ref{tab:zs_CIR_exp_cirr_circo_ori}. 
It is worth noting that existing implementations of the ZS-CIR approach typically rely on the generalization capabilities of VLP-based encoders and are commonly tested with various backbone versions. Therefore, to enable a comprehensive comparison, we list all the performance results of methods utilizing different backbones in our tables.
From these tables, we obtain the following observations. 
1) Certain zero-shot methods yield comparable outcomes to some supervised methods. For example, on the FashionIQ-ori dataset, the top-performing ZS-CIR method, LinCIR (CLIP-G version), attains a score of $55.40$ when averaged across the metrics. Significantly, it not only outperforms all of the traditional encoder-based supervised methods but also manages to achieve a performance comparable to nearly half the number of the VLP encoder-based supervised methods. This implies that even in the absence of manually annotated triplet data, one can still obtain satisfactory retrieval results by ingeniously devising the pre-training strategy and fully activating the potential of powerful VLP capabilities within the CIR context.
2) Typically, the same zero-shot methods, when equipped with a large backbone, consistently deliver better performance. 
For example, LinCIR, ISA, MagicLens, CompoDiff, LDRE, SEIZE, and WieMoCIR all confirm that their larger-scale variants perform more effectively. This further evidences that VLP encoders, which possess well-trained multimodal encoding and cross-modal retrieval capabilities, significantly influence the performance of ZS-CIR.
3) Generally, among the three types of zero-shot methods, the overall performance of methods based on pseudo-triplets is better. For example, on the FashionIQ-ori dataset, regarding the average recall for textual-inversion-based methods, pseudo-triplet-based methods, and training-free methods, the number of methods showing a performance greater than $45.00\%$ is $2$, $6$, and $2$ respectively. On the CIRR dataset, based on R@$10 > 75\%$, the corresponding numbers are $2$, $7$, and $3$. And on the CIRCO dataset, based on mAP@$10 > 30\%$, the numbers are $0$, $3$, and $2$. This can be ascribed to the fact that the pseudo-triplet-based methods still construct triplet data that are most similar to the supervised training paradigm. Although this group of methods is somewhat resource-intensive during the pseudo-triplet construction stage, their overall performance is also the best.

\label{sec:dis}
\section{Discussion and Future Directions}
The introduction of CIR tasks has garnered significant interest due to its requirement for joint reasoning over visual and textual information, making it a challenging endeavor. To better retrieve the target image within the database, the CIR model should comprehend both the image's prototype information and semantic information of modification text. Existing methods in this domain can be broadly categorized into two main groups: supervised CIR and ZS-CIR. We first discuss the issues inherent in each category and point out the possible technological trends of each type. Additionally, we briefly discuss the related tasks of CIR. 

\subsection{Supervised Composed Image Retrieval}
The future work directions for supervised CIR are given as follows.
\begin{itemize}
    \item \textbf{Benchmark Dataset Construction.} As mentioned earlier, most existing CIR datasets are limited to specific domains, \textit{e.g.}, fashion or bird species. Additionally, the false negative issue is prevalent in current datasets due to the data annotation strategy, which first identifies potential image pairs and then annotates the modification text for each pair. Although the open-domain dataset CIRR has been released to address the false negative issue, it still faces two main challenges. First, previous studies~\cite{pic2word,searle} have shown that in many cases, simply relying on the modification text is sufficient to accurately identify the target image in CIRR. Second, CIRR provides only one-to-one query-target correspondence, which does not align with real-world scenarios where multiple images in the dataset may satisfy the retrieval requirements. Creating open-domain CIR datasets with multiple ground-truth target images while mitigating false negatives remains an open challenge requiring further exploration. Furthermore, the scale of existing public datasets is limited; the largest dataset contains only approximately $18.8$M samples. According to the scaling law~\cite{scalinglaw}, the performance and generalization capabilities of CIR improve with larger-scale datasets.  Nevertheless, we must acknowledge that constructing large-scale triplet datasets is both resource-intensive and time-consuming. In fact, MagicLens has made an attempt in this regard by devising a pipeline to generate triplets from web pages in a self-supervised manner, yielding $36.7$M triplets. Unfortunately, this dataset is neither publicly accessible nor sufficiently large. Therefore, we advocate for the development of publicly accessible, large-scale triplet datasets, as such resources would inject new vitality into the field of CIR.
    \item \textbf{LLM-based Image-Text Fusion.} 
    Existing CIR methods have made efforts to devise diverse image-text fusion strategies as reviewed above. While the reasoning capabilities of LLMs or MLLMs have been explored for pseudo triplet generation to address the ZS-CIR task, limited CIR efforts~\cite{zhong2024compositional,mcl} have investigated their potential in encoding input image/text queries and achieving image-text fusion. 
    These approaches use <reference image, modification text, target caption> triplets, generated by their data generation pipelines, for contrastive learning to fine-tune and align the model's encoding capabilities for CIR tasks, while the encoding output is regarded as the fused query embedding. 
    Despite their promising progress, a key issue they suffer from is that improving the encoding capabilities of LLMs may compromise their inherent reasoning abilities. Developing advanced fine-tuning strategies that enhance the encoding capabilities of LLMs while preserving their native reasoning abilities is a critical direction that warrants further investigation.
    \item \textbf{Target Matching with Noisy Data}. 
    Due to the complex nature of the CIR task, manually annotated datasets may contain noisy triplets. For instance, we have empirically observed that some triplets in existing CIR datasets are nonsensical, such as mismatched images and textual descriptions or modification text descriptions unrelated to the corresponding images. Such noisy triplets can mislead the model and hinder its ability to accurately match queries with targets. To address this issue, the model should be capable of learning query-target matching robustly under noisy data conditions. This necessitates the development of adaptive mechanisms that can selectively utilize useful information from the training data while disregarding noise. In conclusion, building a robust CIR model that can effectively handle noisy data remains an open and valuable research direction.
    \item \textbf{Target Matching with Biased Data}. As mentioned earlier, to avoid false negative samples, the annotation of modification text in existing open-domain datasets may sometimes be overly specific. This can lead to scenarios where simply using the text query can correctly retrieve the target image. Consequently, the conventional target matching paradigm may overfit spurious correlations between the text query and the target image while failing to accurately learn the matching relationship between the multimodal query and the target image. This limitation can negatively impact the model's generalization capabilities. To address this issue, the key challenge lies in mitigating the harmful effects while preserving the beneficial contributions of the text query in the context of CIR. Investigating the causal relationships among elements in CIR presents a promising research direction to tackle this challenge.
    \item \textbf{Efficient Reranking.} To enhance the ranking of target images, existing reranking methods in the CIR domain typically conduct more sophisticated point-wise evaluations on the retrieved candidate set, which inherently incurs a high time cost. To ensure that the desired target image can be retrieved, the size of the candidate set for reranking cannot be too small. Consequently, the time cost associated with reranking becomes significant and cannot be overlooked. To make reranking suitable for real-world retrieval scenarios, developing efficient reranking methods is essential and should be a key focus for future research. Furthermore, current reranking and retrieval stages operate independently, with limited correlation. Investigating how these stages can be better integrated to mutually enhance one another represents another promising research avenue.
     \item \textbf{Model Interpretability.} As CIR models grow increasingly complex, their interpretability decreases, making it challenging to understand how decisions are made during the retrieval process. This lack of transparency poses significant challenges for debugging and improving models, as well as for building user trust. Future research directions could focus on designing hybrid models that integrate deep learning with interpretable components or on utilizing post-hoc explanation techniques to shed light on the decision-making process. 
     \item \textbf{Task Synergy.} The triplets in CIR also pertain to other related tasks. For example, text-based image editing~\cite{imageediting1, imageediting2} necessitates a reference image along with a modification text to synthesize the desired target image, which bears a high degree of relevance to the format of CIR. Ideally, integrating retrieval and generation in one framework~\cite{unifashion} would mutually enhance both, achieving a ``kill two birds with one stone'' effect. Specifically, during generation, the model needs to comprehensively envision the desired target image, which would be beneficial for CIR. Moreover, generated low-quality images can be regarded as a type of hard negative, boosting the metric learning in retrieval. Additionally, the retrieval task mainly centers around discriminative features rather than pixel-level generation. Consequently, its complexity is lower and the model converges more readily. This, in turn, can contribute to the continuous learning of the generative process. Moreover, with retrieval capabilities, one can utilize retrieved images as references to attain a retrieval-augmented generation (RAG) effect in the image generation domain. This unified framework also affords users the flexibility to obtain either the retrieved real images or the synthesized ones based on one query. 
\end{itemize}

\subsection{Zero-shot Composed Image Retrieval}
Compared to supervised CIR methods, ZS-CIR approaches eliminate the need for annotated training triplets. Instead, they leverage readily available pre-trained data or utilize modular combinations in a training-free manner to build models. While significant progress has been achieved, several challenges and open questions remain.
\begin{itemize}
    \item \textbf{Pseudo Triplets Generation.}
    Some researchers have explored the automatic generation of pseudo triplets, enabling CIR model training without the need for manually annotated triplets. However, current methods typically use an existing captioning model to convert an image into a caption and then generate the modification text with LLMs. These approaches may suffer from information loss during the modality transformation, as the generated image captions often capture only coarse-grained attributes (\textit{e.g.}, the color of a dress) while missing fine-grained visual details (\textit{e.g.}, the logo on the dress). This limitation has two main adverse effects: it constrains the identification of potential reference-target image pairs and results in coarse-grained modifications, failing to capture more challenging fine-grained variations. Furthermore, using uniform prompts to generate all triplets can reduce the diversity of generated samples. In contrast, real-world users often express modification requirements in various styles. As a result, building high-quality pseudo triplets for training CIR models remains a significant challenge.
 \item \textbf{MLLM-based Training-free Methods.} Without additional design of pre-training tasks, training-free methods present themselves as straightforward and promising options in this field. Typically, existing training-free methods predominantly utilize LLMs to summarize the caption of the reference image along with the modification text, thereby transforming the task into text-to-image retrieval. However, relying on second-hand information on the reference image might lead to bottlenecks and only attain suboptimal performance. With the rapid development of MLLMs, a promising research avenue is to explore the direct utilization of MLLMs to process multimodal queries for ZS-CIR.
    \item \textbf{Efficient Retrieval.} The current evaluation protocol primarily concentrates on retrieval efficacy while neglecting retrieval efficiency. Consequently, the devised methods merely focus on enhancing retrieval performance. However, such approaches might not be practical in real-world scenarios. For example, certain methods~\cite{ldre, seize} make LLM reason about the multimodal query multiple times to enhance the retrieval effect, yet this incurs a high time cost. Therefore, striking a balance between efficacy and efficiency is crucial to ensure that the research not only remains relevant in academia but also holds potential application value. This entails two aspects. Firstly, the method itself ought to be lightweight. Secondly, during the similarity computing process, some accelerated retrieval techniques, such as hash learning~\cite{hash} and index structure~\cite{index}, should also be explored.
   \item \textbf{Few-shot CIR.} The generalization ability of existing supervised CIR models is often limited by the small scale of available datasets, while ZS-CIR completely avoids using labeled triplets. In practice, however, annotating a small number of samples is often feasible. Consequently, some pioneering studies~\cite{wu2023few,hou2024pseudo} have started to investigate the few-shot CIR task, which leverages a limited number of annotated triplets to enhance model training. A key research question in this area is how to select high-quality samples for annotation to maximize the benefits of manual annotation efforts.


\end{itemize}

\subsection{Relation Tasks of Composed Image Retrieval}
For the related tasks of CIR, current methods remain underdeveloped and require further refinement to fully leverage the unique attributes of each specific sub-task. This indicates a substantial opportunity for improvement by tailoring approaches more closely to the distinct characteristics and requirements of these tasks. 
In addition to refining current methodologies, exploring a broader range of novel and innovative sub-tasks presents an exciting avenue for future research. Expanding CIR tasks could lead to groundbreaking advancements in the field, opening up new possibilities for application and enhancing our understanding of CIR as a whole. Future research should focus on developing sophisticated models that are capable of seamlessly incorporating diverse modalities, thereby enriching the retrieval process and improving the accuracy and relevance of results across a variety of contexts.

\section{Conclusions}
\label{sec:concl}
In this paper, we provide a comprehensive review of state-of-the-art methods in CIR. Specifically, we begin by reviewing supervised CIR approaches, with a focus on the four main components of their general framework. We then explore the emerging field of zero-shot CIR, along with several related tasks in CIR. Following this, we review various CIR datasets, grouping results according to the datasets and providing corresponding analyses. 
Through these comprehensive endeavors, we have gained profound insights into the current challenges in CIR and have outlined some promising research directions for the future. This comprehensive survey is an invaluable and insightful resource for researchers and practitioners, significantly propelling the rapidly evolving field of CIR.


\begin{acks}
This work is supported by the National Natural Science Foundation of China under Grant 62376137 and 62206157; Natural Science Foundation of Shandong Province under Grant ZR2022YQ59 and ZR2022QF047.
\end{acks}

\bibliographystyle{ACM-Reference-Format}
\bibliography{sample-base}



\end{document}